\documentclass{aa}
\usepackage{epsfig}
\usepackage{natbib}
\usepackage{graphics}
\usepackage{txfonts}
\usepackage{color} 
\usepackage{ulem}
\newfont{\tss}{cmssdc10 scaled 950}

\begin{document}

\title{A nearby GRB host galaxy: VLT/X-shooter observations of \\
  HG 031203\thanks{Based on 
observations collected at the European Southern Observatory, Chile, ESO 
programme 60.A-9024(A).}}

\author{N. G.\ Guseva \inst{1,2}    
\and Y. I.\ Izotov \inst{1,2}
\and K. J.\ Fricke \inst{1,3}
\and C.\ Henkel \inst{1}}
\offprints{N. G. Guseva, guseva@mao.kiev.ua}
\institute{          Max-Planck-Institut f\"ur Radioastronomie, Auf dem H\"ugel 
                     69, 53121 Bonn, Germany
\and
                     Main Astronomical Observatory,
                     Ukrainian National Academy of Sciences,
                     Zabolotnoho 27, Kyiv 03680,  Ukraine
\and
                     Institut f\"ur Astrophysik, 
                     G\"ottingen Universit\"at, Friedrich-Hund-Platz 1, 
                     37077 G\"ottingen, Germany
}
\date{Received \hskip 2cm; Accepted}

\abstract
{Long-duration gamma-ray bursts (LGRBs), which release enormous amounts
of energy into the interstellar medium, occur in
galaxies of generally low 
metallicity. For a better understanding of this phenomenon, detailed 
observations of the specific properties of the host galaxies (HG) and the 
environment near the LGRBs are mandatory.
} 
{We aim at a spectroscopic analysis of HG 031203, the host galaxy of a 
LRGB burst, to obtain its properties. Our results will be compared with those 
of previous studies and the properties of a sample of luminous compact 
emission-line galaxies (LCGs) selected from SDSS DR7.
}  
{Based on VLT/X-shooter spectroscopic observations
taken from commissioning mode in the 
wavelength range $\sim$$\lambda\lambda$3200 -- 24000\AA, we use
standard direct methods to evaluate physical conditions and 
element abundances.
The resolving power of the instrument also allowed us to trace
the kinematics of the ionised gas. 
Furthermore, we use X-shooter data together with
{{\sl Spitzer}} observations in the mid-infrared range for testing
hidden star formation.
} 
{We derive an interstellar oxygen abundance 
of 12 + log O/H = 8.20 $\pm$ 0.03 for HG 031203.
The observed fluxes of hydrogen lines 
correspond to the theoretical recombination values after correction for  
extinction with a single value $C$(H$\beta$) = 1.67. We produce
the CLOUDY photoionisation H {{\sc ii}} region model that reproduces observed 
emission-line fluxes of different ions in the optical range. This model
also predicts emission-line fluxes in the near-infrared (NIR) and mid-infrared 
(MIR) ranges that agree well with the observed ones. This implies that 
the star-forming region observed in the optical range is the only source of 
ionisation and there is no additional source of ionisation seen in the 
NIR and MIR ranges that is hidden in the optical range.
We find the composite kinematic structure from profiles of the strong emission
lines by decomposing them into two Gaussian narrow and broad components. 
These components correspond to two H {{\sc ii}} regions, separated by 
$\sim$34 km s$^{-1}$, and have full widths at half maximum 
(FWHM) $\sim$115 and $\sim$270 km s$^{-1}$, respectively.
 We find that the heavy element abundances, 
extinction-corrected H$\alpha$ luminosity  
$L$(H$\alpha$)=7.27 $\times$ 10$^{41}$ erg s$^{-1}$, 
stellar mass $M_*$ = 2.5$\times$10$^8$$M_{\odot}$, star-formation rate 
SFR(H$\alpha$) = 5.74 $M_\odot$ yr$^{-1}$ and specific star-formation 
rate SSFR(H$\alpha$) = 2.3$\times$10$^{-8}$ yr$^{-1}$ of HG 031203
are in the range that is covered by the LCGs.
This implies that the LCGs with extreme star-formation 
that also comprise green pea galaxies as a subclass may harbour GRBs.
}
{}
\titlerunning{A nearby GRB 031203 host galaxy based on VLT/X-shooter observations}
\authorrunning{N.G.Guseva et al.}
\maketitle


\section{Introduction \label{intro}}

GRB 031203 is one of the two closest ($z$ $<$ 0.2) 
long-duration gamma-ray bursts 
(LGRB) known apart from the exceptional GRB 980425. 

Most soft-spectrum LGRBs are accompanied by massive stellar 
explosions \citep[the GRB-supernovae (SNe) connection, see][] 
{Paradijs1999,Woosley2006,Soderberg2006,HjorthBloom2011}.
  This and other evidence links LGRBs to ongoing star formation  
and consequently to the most massive stars as possible progenitors of LGRBs 
and X-ray flares (XRFs) 
\citep{Bloom2002,LeFloch2003,Christensen2004,Fruchter2006}.
Wolf-Rayet (WR) spectral features were found in the spectra of several LGRB 
host galaxies (HGs) \citep{Hammer2006,Han2010}.
\citet{Savaglio2009} present an extensive study of the 
most extensive sample of HGs to date, encompassing 46 
targets. The authors conclude that there is no 
compelling evidence that HGs are peculiar galaxies and
that they are instead similar to normal star-forming galaxies in the 
local and distant universe.  Accordingly, HGs are proposed to be used as 
tracers of star formation.
  Spectroscopic studies of HGs give information on LGRB progenitors
and on the physical properties of regions hosting LGRBs. 
  Investigations of HGs have been performed for many galaxies
\citep{Christensen2004,Prochaska2004,Stanek2006,Sollerman2005,Wiersema2007,Margutti2007,Thoene2007,Hammer2006,Kewley2007,Thoene2008,Christensen2008,Savaglio2009,Han2010,Svensson2010,Watson2010,Levesque2010,Levesque2010b,Levesque2011,Schady2010,Vergani2011,Wiersema2011}.
 However, to date spatially resolved studies have been performed only for 
several low-redshift HGs 
\citep{Hammer2006,Thoene2008,Christensen2008,Sollerman2005,Levesque2011}.
 For more distant emission-line galaxies at redshifts $z$ $\ga$ 0.1, 
the properties of the LGRB environment can be retrieved only from the 
interstellar medium (ISM) of the entire galaxy, as is the case for 
HG 031203.

\begin{figure}[t]
\hspace*{0.0cm}\psfig{figure=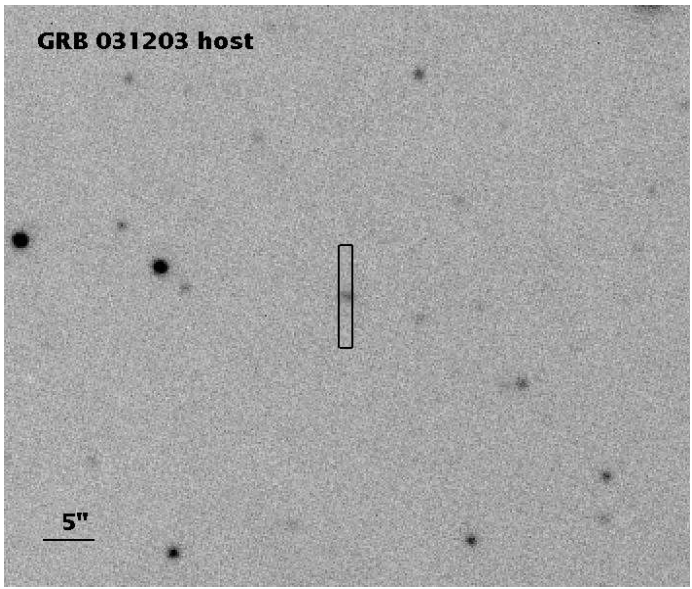,angle=0,width=9.cm,clip=}
\caption{X-shooter acquisition image of HG 031203
[ESO program 60.A-9442(A)] 
showing the slit location. The galaxy has a compact, slightly 
elongated shape and likely consists of two H {\sc ii} regions.
North is up and East is to the left.
}
\label{fig0}
\end{figure}

\begin{figure*}[t]
\hspace*{0.5cm}\psfig{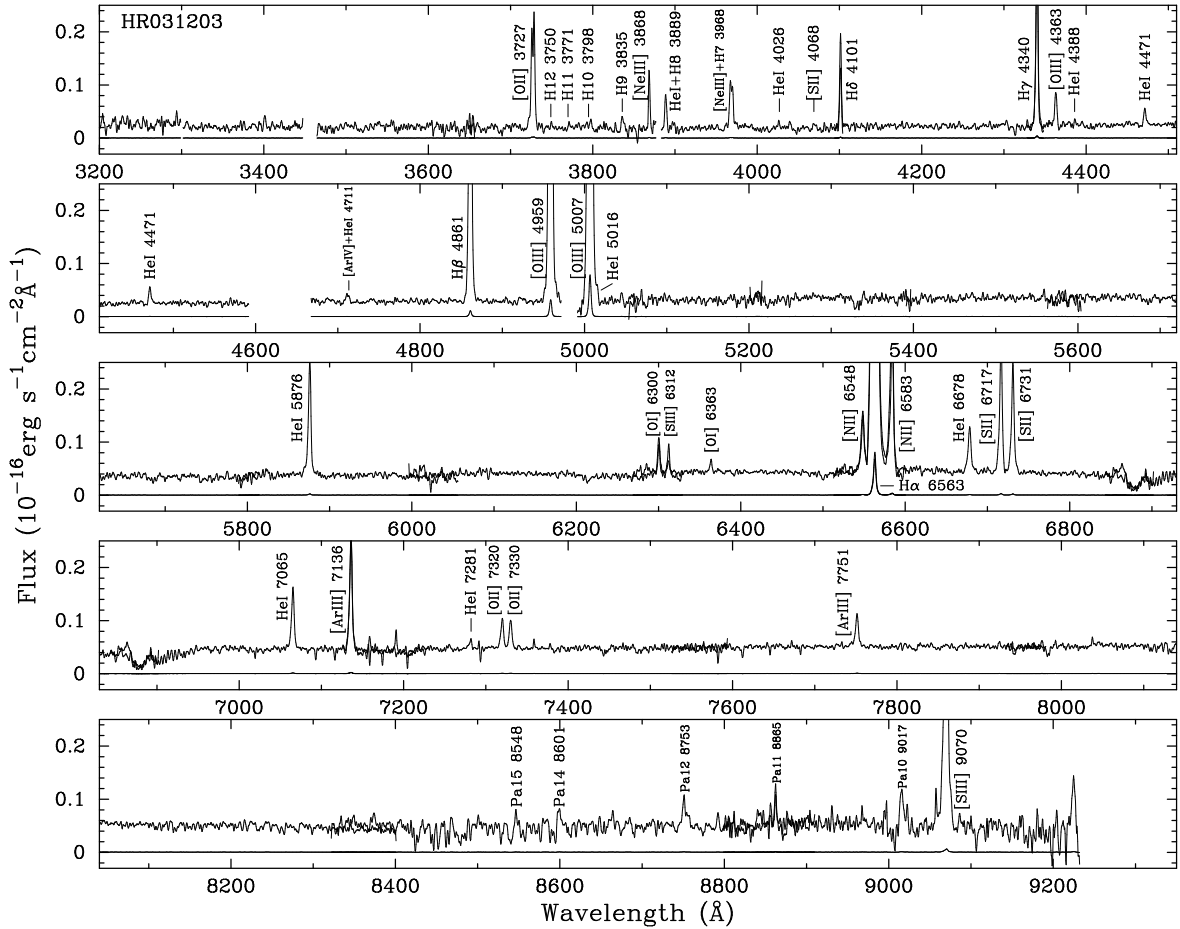}
\caption{Flux-calibrated VLT/X-shooter near-UV and optical range 
(UVB + VIS arms) spectrum 
of the GRB 031203 host galaxy
corrected for a redshift 
of $z$ = 0.1055 
(upper spectrum in each panel).
The lower spectrum is the upper spectrum downscaled by a factor of 100.
The scale of the ordinate is that of the upper spectrum.
}
\label{fig1}
\end{figure*}

\begin{figure*}[t]
\hspace*{0.5cm}\psfig{figure=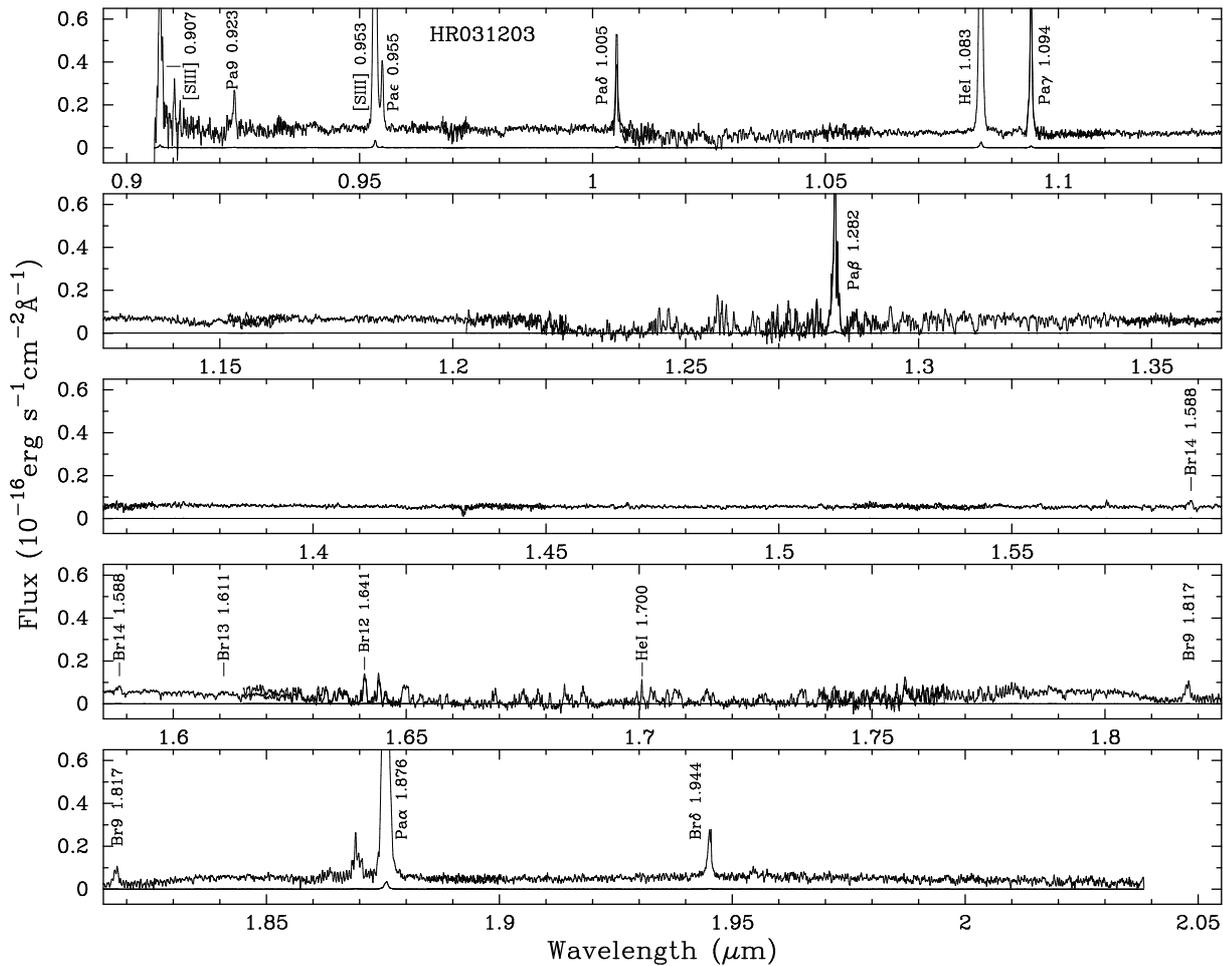,angle=-90,width=16.cm,clip=}
\caption{Same as Fig. \ref{fig1} but for the near-infrared (NIR) range.
}
\label{fig2}
\end{figure*}

\citet{Svensson2010}, \citet{Levesque2010}, \citet{Savaglio2009},
\citet{Kewley2007} and \citet{Wiersema2007} 
found that HGs are relatively low-mass
galaxies with a low-metallicity ISM and intense star formation.
For a given mass (or luminosity) HGs are systematically
offset towards lower metallicities in  
luminosity-metallicity ($L-Z$) \citep{Margutti2007,Stanek2006,Kewley2007,Levesque2010}
or mass-metallicity ($M-Z$) \citep{Savaglio2009,Svensson2010,Levesque2010b,Vergani2011} 
relations as compared to dwarf irregulars and normal star-forming 
emission-line galaxies 
\citep{Lequeux1979,Skillman1989,RicherMcC1995,KobylZarit1999,MelbourneSalzer2002,Lee2004,Pil2004,Kong2004,Shi2005,Lee45mu2006}.
 Moreover,  \citet{Levesque2010b} obtained the offset between nearby HGs 
with $z$ $<$ 0.3  and nearby
SDSS star-forming galaxies as well as between HGs with 
intermediate-redshift
(0.3 $<$ $z$ $<$ 1) and emission-line galaxies from the DEEP2 survey 
($<$$z$$>$=0.8). It was also established that LGRB HGs have lower 
metallicities than SN HGs without accompanying LGRBs
\citep{Fruchter2006,Modjaz2008,Levesque2010}.
\citet{Kobulnicky2003}  and \citet{Savaglio2005} have found a temporal 
evolution of emission-line galaxies in $L-Z$ and $M-Z$ diagrams, 
respectively. Higher 
redshift galaxies are more metal-poor at a fixed mass or luminosity 
than local emission-line galaxies.
Host galaxies (HGs) are characterised by high star-formation rates 
(SFRs) and high specific star-formation rates (SSFRs) 
\citep{Savaglio2009,Svensson2010,Levesque2010,Levesque2010b}.
  
The metallicities of most HGs are obtained using empirical 
strong-line methods, such as $R$23, O3N2, mass-metallicity relation 
and other calibrations 
\citep{Savaglio2005,Savaglio2009,Levesque2010,Levesque2010b,Svensson2010,Vergani2011} and are somewhat uncertain. This is because of 
a well-known offset between the metallicity obtained by the 
direct $T_e$-method and empirical strong-line methods. Oxygen abundances 
obtained by empirical methods are by $\sim$ 0.2--0.6 dex higher than those 
obtained with the $T_e$-method \citep{G2009,Hoyos2005,Shi2005}. 
Moreover, the luminosity-metallicity relation is also blurred because the 
metallicity determinations of various galaxy samples do not employ a 
unique technique. Therefore, we use only a comparison sample of 
HGs for which the metallicity is derived using the $T_e$-method. 
  
The metallicity of the ISM in the region around the LGRB site
may be different from that averaged over the whole galaxy. 
However, oxygen abundance variations over dwarf compact galaxies are small 
($\sim$0.1 dex), that is, in the range of 
errors of abundance determinations \citep[see, e.g. ][]{Papaderos2006,I06b}.
 Therefore, the average metallicity of the dwarf HG is comparable to that
of the LGRB environment. Thus, in low-mass galaxies the metallicity
of an entire galaxy can be a proxy of an LGRB environment.

 Therefore, a comprehensive study of HGs that exhibit 
bright H {\sc ii} region features is essential for a better
 understanding of
LGRBs and needs to investigate the physical conditions, chemical 
abundances, extinction, and kinematic structure of their environment.

The long-duration ($\sim$30 s) GRB 031203 was discovered in 2003 with 
INTEGRAL \citep{Gotz2003}.
 A compact dwarf galaxy coinciding with the X-ray source \citep{Hsia2003}
was later identified as the GRB host HG 031203 at redshift $z$ = 0.1055
\citep{Prochaska2004}. 
  Monitoring of the galaxy to search for 
SNe events led to the discovery of SN2003lw \citep{Tagliaferri2004}.
The proximity of the galaxy (it is one of the closest known long-duration GRB 
host galaxies) allows us to perform a detailed analysis of its properties.

An oxygen abundance 12 + log O/H = 7.96 ($T_e$-method) and high 
reddening $E$($B-V$) = 1.17 corresponding to $C$(H$\beta$) = 1.72 were 
obtained by \citet{Levesque2010} from Keck I LRIS observations.
Based on optical VLT spectra \citet{Margutti2007} derived an 
oxygen abundance 12 + log O/H = 8.12 $\pm$ 0.04.
 \citet{Prochaska2004} from Magellan/IMACS optical spectroscopy 
obtained an oxygen abundance 12 + log O/H = 8.02 $\pm$ 0.15.
This galaxy is located at low galactic latitude, therefore the
extinction by the Milky Way is high. 
Special attention is needed to derive 
background and intrinsic extinction in HG 031203,
which can affect the abundance determination.
  \citet{Margutti2007} and \citet{Prochaska2004} analysed both 
the Milky Way and the intrinsic extinction 
in HG 031203 and derived intrinsic extinction coefficients 
$C$(H$\beta$) = 0.59 and $C$(H$\beta$) = 0.46, respectively. 
  Their intrinsic host-galaxy extinctions seem higher that usually obtained
in star-forming compact dwarf galaxies. 
 This is mainly because the authors assumed a lower 
value of the Galactic reddening than that given by \citet{Schlegel1998}. 
On the other hand, the available literature data 
show no evidence for substantial extinction in any comprehensively 
studied HGs \citep{Sollerman2005,Savaglio2009}. Nine out 
of eleven well-studied GRB HGs collected by \citet{Savaglio2009} have 
$A$($V$) in the range $\sim$0--0.6 [or $C$(H$\beta$) $\sim$0--0.3] as 
derived from the Balmer decrement. For HG 030329 and HG 980425 
\citet{Sollerman2005} derived $C$(H$\beta$)$_{\rm HG}$ of 0.06 and 0.10, 
respectively.    
  Therefore, it is likely that \citet{Margutti2007} and \citet{Prochaska2004}
overestimated the intrinsic reddening in HG 031203.

\begin{figure}[t]
\hspace*{0.0cm}\psfig{figure=aa16765-11f4.ps,angle=-90,width=9.cm,clip=}
\caption{Baldwin-Phillips-Terlevich (BPT) diagram \citep{BPT1981}
for luminous compact galaxies (LCGs) from SDSS DR7
\citep{IGT2010} (small blue circles).   
HG 031203 (this paper) is shown by a large filled red star. 
The five low-metallicity AGNs from \citet{IT08} and  \citet{I10} are 
indicated by large filled black circles. 
Host galaxy (HG) 031203 for any data is shown by a star. 
 Other HGs are denoted as follows (see text for details):  
\citet{Savaglio2009}: filled green squares,
\citet{Levesque2010}: filled light blue squares,
\citep{Christensen2008}: violet asterisks, 
 \citet{Hammer2006}: large green open circles,
\citep{Wiersema2007}: filled purple triangle, 
\citet{Han2010}: purple diamonds,
\citet{Margutti2007}: filled purple star,
\citet{Prochaska2004}: open blue star,
\citet{Watson2010}: open yellow star.
   All HGs with emission line fluxes lower than
2 $\times$ 10$^{-17}$ erg s$^{-1}$ cm$^{-2}$ are shown 
by small symbols.
Also, the100,000 emission-line galaxies from SDSS DR7 selected by 
\citet{IGT2010} are shown by grey dots. 
  The dashed line represents the empirical line of \citet{Kauff2003} 
that separates galaxies dominated either by star formation or by an
AGN. The continuous line shows the upper 
limit for purely star-forming galaxies (SFGs) from
\citet{Stasinska2006}. 
\vspace{0.1cm}
\hspace{10.5cm} (A colour version of this figure is available in the online journal.)
}
\label{diag}
\end{figure}

\begin{figure}[t]
\hspace*{0.0cm}\psfig{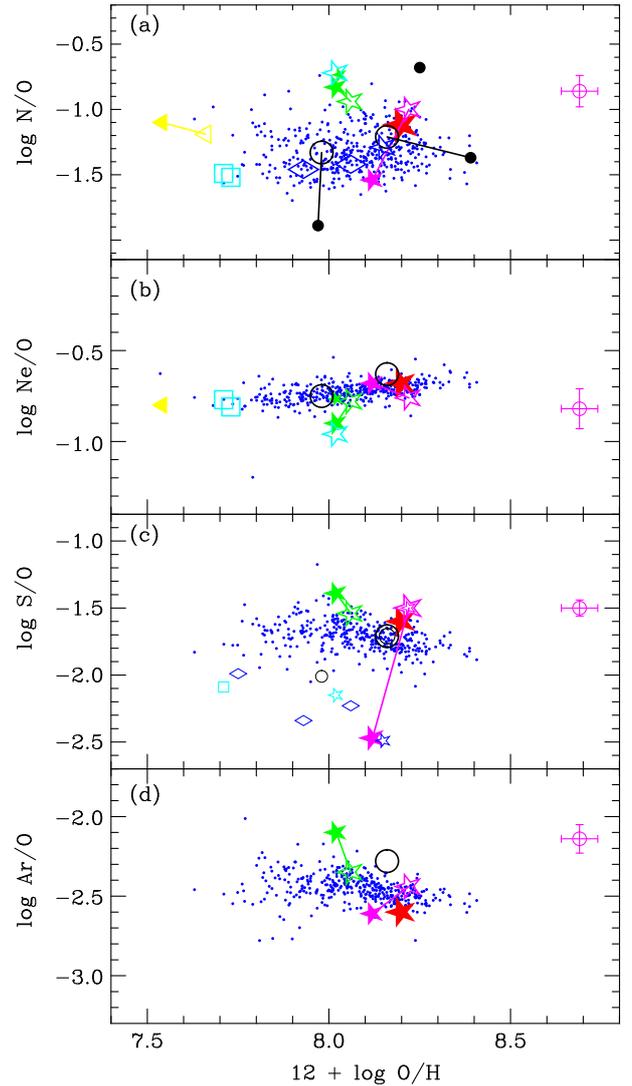}
\caption{Abundance ratios log N/O (a), log Ne/O (b), log S/O (c) and
log Ar/O (d) vs. oxygen abundance 12 + log O/H  
for HG 031203 (this paper) and for LCGs from 
SDSS DR7 \citep{IGT2010} are shown by 
large red filled stars and
small blue filled circles, respectively.  
 Data from the literature and recalculated from published emission line fluxes
are shown by filled and open symbols, respectively. 
All data for HG 031203 are shown by stars of different colours. 
Specifically, \citet{Margutti2007}: filled and open purple stars, 
\citet{Hammer2006}: filled and open black circles, 
\citet{Wiersema2007}: filled yellow triangles, 
\citet{Levesque2010}: open light blue squares, 
\citet{Han2010}: open blue rhombs, 
\citet{Prochaska2004}: filled green stars (see text).
The solar abundance ratios by \citet{Asplund2009} are indicated by the large 
purple open circles and the associated error bars. 
\vspace{0.1cm}
\hspace{10.5cm} (A colour version of this figure is available in the online journal.)
}
\label{oiioiii}
\end{figure}

We here present high-quality archival  VLT/X-shooter spectroscopic observations
of HG 031203 over a wide wavelength range, which allows us
to derive physical conditions, intrinsic 
reddening and element abundances in the ionised 
interstellar medium of the galaxy.
  The same observations were discussed by \citet{Watson2010}. 
However, they did not draw any conclusions on extinction and heavy 
element abundances.
 The observations encompass the  
near-infrared (NIR) range that allows us to find out whether
hidden star formation is present in this galaxy. 
Such a broad wavelength range extending to the NIR range permits us to
derive a more reliable stellar mass (and consequently SSFR)
of the galaxy. This is because cool low-mass stars mainly contribute
to the stellar mass and they also emit mainly in the NIR range. 
  We note, however, that in low-metallicity compact dwarf 
galaxies with high SFR
the presence of strong ionised gaseous emission complicates the stellar mass
determination. We also use the {\sl Spitzer} observations of
\citet{Watson2010} for testing more hidden star formation in the mid-infrared
(MIR) range.

The observations are described in Sect.~\ref{obs}. 
In Sect.~\ref{results} we present the main properties of the host galaxy.
 In particular, the diagnostic diagram is given in subsect.~\ref{diagn}. 
 The element abundances in HG 031203
are presented in  subsect.~\ref{abund}. 
 The extinction and hidden star formation  
are discussed in subsect.~\ref{extinc},
the CLOUDY modelling and the comparison of the observed and predicted
emission-line fluxes in the MIR range are discussed in subsect.~\ref{CLOUDY},
the kinematic structure in subsect.~\ref{kinem},
the luminosity-metallicity relation in 
subsect.~\ref{LZr} and the  
star-formation rate in subsect.~\ref{SFR}.
Our conclusions are summarised in Sect.~\ref{concl}.

\section{Observations \label{obs}}

A new spectrum of the GRB 031203 host galaxy was obtained 
during the commissioning of the VLT X-shooter on 2009 March 17
[ESO program 60.A-9024(A)]. 
In Fig.~~\ref{fig0} the acquisition image 
of HG 031203 with the slit location is shown.
  The galaxy presents itself as a slightly elongated source with major axis 
$\sim$ 1\arcsec, which is 
comparable to the slit width of 0\farcs9 -- 1\arcsec. 
Two H {\sc ii} regions are present in the galaxy, one brighter than the other.
   The observations were performed in the 
wavelength range $\sim$$\lambda$3200 -- 24000\AA\ at the airmass  1.037
using the three-arm echelle X-shooter spectrograph mounted at the UT2 
Cassegrain focus.
In the UVB, VIS and NIR arms the total exposure time 
of 4800s were broken into four equal subexposures of 1200s each.
  Nodding along
the slit was performed according to the scheme ABBA with the object positions
A or B differing by 5\arcsec\ along the slit. 
 In the UVB arm with wavelength range $\lambda$3233 -- 5600\AA\
a slit of 1\arcsec $\times$ 11\arcsec\ was used.
  In the VIS and NIR arms with wavelength ranges $\lambda$5475 -- 10206\AA\
and $\lambda$11100 -- 23985\AA, respectively, a slit of 
0\farcs9 $\times$ 11\arcsec\ were used.
The binning factor along the spatial and dispersion axes in the UVB and
VIS arms was 1, except for the UVB arm where the binning factor along the 
dispersion axis was 2.
  The above instrumental set-up resulted in resolving powers 
$\lambda$/$\Delta$$\lambda$ of 10200, 8800, and 5100 for the UVB,
VIS, and NIR arms, respectively. The seeing was $\sim$ 1\arcsec.
All observations were obtained at low airmass of $\sim$ 1.03 - 1.06, 
therefore the effect of the atmospheric dispersion is low. The 
correction for this effect was applied to the UBV and VIS arm spectra during
observations.
 We used the spectrum of the spectrophotometric standard star
GD 153 for the flux calibration. The absolute synthetic fluxes for this 
star were taken from the Space Telescope Science Institute web site
\footnote{ftp://ftp.stsci.edu/cdbs/current$_-$calspec/}.
 Spectra of thorium-argon (Th-Ar) comparison arcs were 
 used for wavelength calibration of the UVB and VIS arm 
observations. For the wavelength calibration of the NIR spectrum we used 
night-sky emission lines.

In a first step cosmic ray hits of all UVB, 
VIS, and NIR spectra were removed using the routine 
CRMEDIAN. The remaining
hits were subsequently removed manually after background subtraction.
  The two-dimensional UVB and VIS spectra were bias-subtracted and 
flat-field corrected using IRAF\footnote{IRAF is 
the Image Reduction and Analysis Facility distributed by the 
National Optical Astronomy Observatory, which is operated by the 
Association of Universities for Research in Astronomy (AURA) under 
cooperative agreement with the National Science Foundation (NSF).}. 
The two-dimensional NIR spectra were corrected for dark current and 
divided by the flat frames to correct for the pixel 
sensitivity variations.
 For each of the UVB, VIS, and NIR arms we separately coadded spectra
with the object at the position A and spectra with the object 
at the position B. 
Then, the coadded spectrum at the position B was subtracted from the
coadded spectrum at the position A. This resulted in a frame with
subtracted background. We used the IRAF
software routines IDENTIFY, REIDENTIFY, FITCOORD, and TRANSFORM to 
perform wavelength
calibration and correct for distortion and tilt for each frame. 
The one-dimensional wavelength-calibrated spectra were then extracted from the 
two-dimensional frames using the APALL routine. 
We adopted extraction apertures of  
1\arcsec $\times$ 2\farcs5, 0\farcs9 $\times$ 2\farcs5 and 
0\farcs9 $\times$ 2\farcs5 for the UVB, VIS and NIR spectra, respectively.
Before extraction, the spectra at the positions A and B in the two-dimensional
background-subtracted frames were carefully aligned with the routine ROTATE 
and co-added. For the flux calibration we adopted an atmospheric 
extinction curve for the Cerro Tololo Observatory, which is very similar to
that for Cerro Paranal \citep{P11}. Although telluric standards were 
also observed, no attempts were made to correct for telluric absorption.
This is because this correction resulted in very noisy spectra in the 
regions of strong absorption and does not help much in the determination of
the reliable fluxes of the lines that fall in those wavelength ranges. 
Fortunately, there are many strong lines in regions of low or no telluric 
absorption. The number of these lines is sufficient for our analysis.
In particular, the redshifted strongest NIR emission line Pa$\alpha$ is in
the region without telluric absorption.
  The resulting flux-calibrated continuum in all orders
is monotonic over the entire range of the spectrum. Therefore,
no adjustment of the UVB, VIS, and NIR spectra were needed. 

The resulting flux-calibrated and redshift-corrected UVB and VIS spectra of 
HG 031203 are shown in Fig.~\ref{fig1} and the resulting flux-calibrated 
and redshift-corrected NIR spectrum is shown in Fig.~\ref{fig2}.

\section{Results \label{results}}

\subsection{Diagnostic diagram\label{diagn}}

The position of HG 031203 in the 
[O {\sc iii}] $\lambda$5007/H$\beta$ vs.
[N {\sc ii}] $\lambda$6583/H$\alpha$ 
Baldwin-Phillips-Terlevich (BPT) diagram \citep{BPT1981}
is shown in Fig.~\ref{diag} by a large filled red star.
Luminous compact emission-line galaxies (LCGs) from SDSS DR7 
\citep{IGT2010} are shown by small blue circles.
The global properties of LCGs are similar to those of the 
green pea galaxies, which were
recently discovered by \citet{Cardamone2009} as a new class of luminous
compact galaxies at redshift $z$ $\sim$ 0.1 -- 0.3 with a peculiar bright
green colour.
 Based on the photometrically selected sample of 251 galaxies, 
\citet{Cardamone2009} found that green pea galaxies have some of the highest
specific star-forming rates seen in the local Universe and are similar in size,
 mass, luminosity, and metallicity to luminous blue compact galaxies and 
to high-redshift ultraviolet luminous galaxies (Lyman-break galaxies and 
Ly-$\alpha$ emitters).
\citet{IGT2010} constructed and studied an extensive sample of 803 
star-forming LCGs in a wider redshift range 
$z$ = 0.02 -- 0.63, 
selected from SDSS DR7 with the use of spectroscopic and photometric criteria. 
The five low-metallicity AGN candidates from \citet{IT08} and  
\citet{I10} are shown by large filled black circles. 
   Data of other authors for HG 031203 are shown by stars
in different colours. 
 Other HGs are denoted as follows:  
\citet{Savaglio2009}: filled green squares (031203, 020903 and 060505),
\citet{Levesque2010}: filled light blue squares (030329, 020903 and 060218),
\citep{Christensen2008}: violet asterisks (980425 -- WR, SN regions and 
entire host galaxy), 
 \citet{Hammer2006}: large green open circles (020903 and 980425 (WR and 
SN regions)),
\citep{Wiersema2007}: filled purple triangle (060218), 
\citet{Han2010}: purple diamonds (020903, 030329, 031203, 060218, 060505 GRB 
site and entire galaxy),
\citet{Margutti2007}: filled purple star (031203),
\citet{Prochaska2004}: open blue star (031203),
\citet{Watson2010}: open yellow star (031203).
   All HGs with emission line fluxes lower than
2 $\times$ 10$^{-17}$ erg s$^{-1}$ cm$^{-2}$ are shown by small symbols.
Also, the 100,000 emission-line galaxies from SDSS DR7  
are seen as a cloud of grey dots.
  The dashed line represents the empirical line of \citet{Kauff2003}, 
which separates  star-forming galaxies and AGNs. 
The continuous line delineates the 
upper limit for pure star-forming galaxies from \citet{Stasinska2006}.

\begin{figure*}
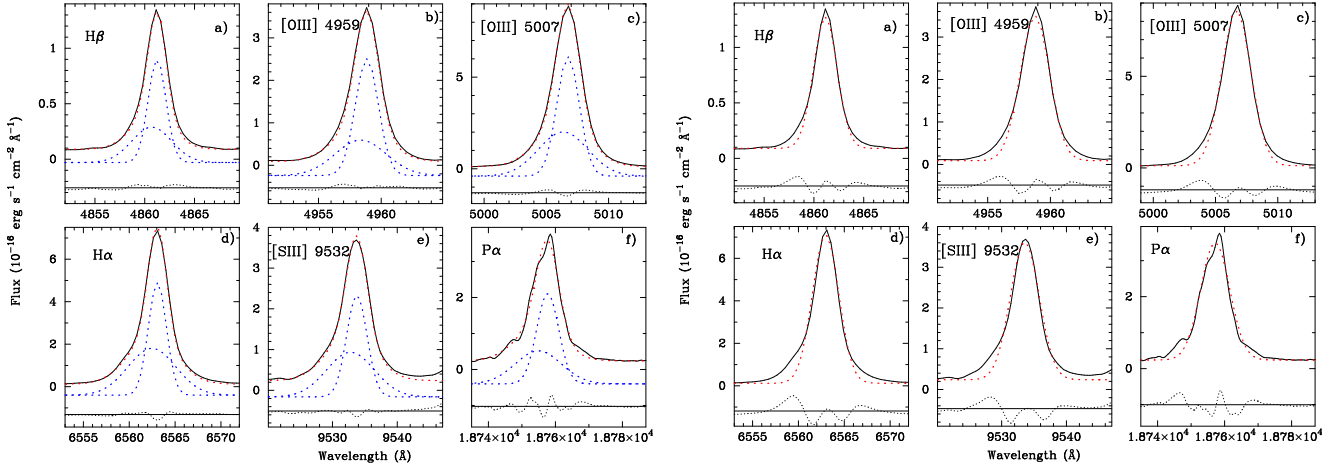

\hspace*{0.0cm}\psfig{figure=aa16765-11f6a.ps,angle=-90,width=8.5cm}
\hspace*{0.3cm}\psfig{figure=aa16765-11f6b.ps,angle=-90,width=8.5cm}
\caption{\textit{Left panel.}
Decomposition of strong emission-line profiles into two Gaussian 
components in the spectrum of HG 031203 for:
a) H$\beta$ $\lambda$4861; b) [O {\sc iii}] $\lambda$4959; 
c) [O {\sc iii}] $\lambda$5007; d) H$\alpha$ $\lambda$ 6563; 
e) [S {\sc iii}] $\lambda$9532
and f) Pa$\alpha$ $\lambda$18756.
The observed spectrum and the fit are shown by black solid and red 
dashed lines. The two Gaussian components and residual spectra
are shown by blue dashed and black dotted lines. 
For convenience the observed spectra, Gaussians, and residual are shifted. 
\textit{Right panel.} The same as in \textit{left panel}, but 
the strong emission line profiles fitted by the single Gaussian.
}
\label{profiles}
\end{figure*}

\begin{figure*}[t]
\hspace*{0.5cm}\psfig{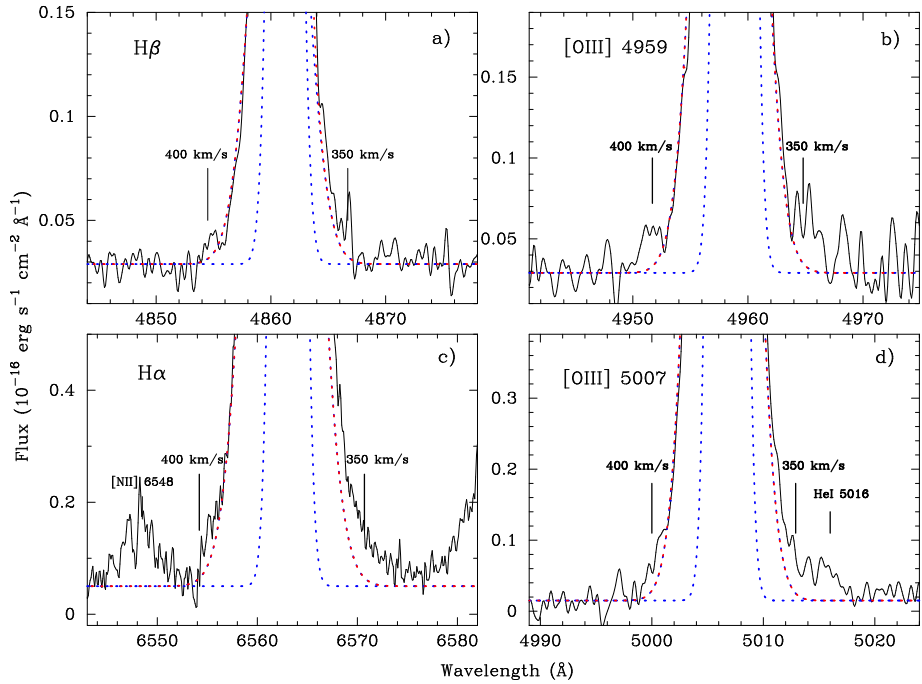}
\caption{Velocity excess in emission line profiles of 
a) H$\beta$ $\lambda$4861; b) [O {\sc iii}] $\lambda$4959;
c) H$\alpha$ $\lambda$ 6563 and d) [O {\sc iii}] $\lambda$5007.
Observational data are shown by solid lines and fits by dotted lines 
(narrow and broad components of lines in blue and summed fits in red).
 The positions of the blue-shifted $v$=400 km s$^{-1}$ and
red-shifted $v$=350 km s$^{-1}$ broad components 
are shown by straight lines. 
}
\label{prof}
\end{figure*} 

In Fig.~\ref{diag} almost all LCGs lie in the low-metallicity 
part of what is usually considered as the region of star-forming galaxies. 
HG 031203 is situated on the border between SFG and 
AGN regions and is close to the region of very low-metallicity AGNs.
Photoionisation
models of AGN show that lowering their metallicity 
moves them to the left of BPT diagram, so that they end up in the SFG 
region \citep{Stasinska2006}.
  \citet{IT08} have concluded that the BPT diagram is unable to distinguish 
between SFGs and low-metallicity AGNs. 
   The five low-metallicity AGN candidates \citep{IT08,I10} 
also contain the broad components of strong emission lines H$\alpha$ and
H$\beta$ with high luminosities and very steep Balmer decrement.  
  The position of AGN candidates in BPT diagram
can be accounted for by the models with the non-thermal radiation 
contributing less than $\sim$ 10\% to the total radiation \citep{IT08}.
In particular, the pure photoionisation model without nonthermal radiation 
agrees quite well with observations (see subsect. 3.4 in this paper).
 \citet{Watson2010} rule out the presence of an AGN in HG 031203
based on flux ratios of optical emission lines (X-shooter data) 
and mid-IR spectrum.
The authors confirm the previous conclusions of \citet{Prochaska2004} 
and \citet{Margutti2007}.
On the other hand, \citet{Levesque2010}, using line-flux ratios 
from the Keck I LRIS spectrum, concluded that HG 031203 shows
evidence of AGN activity.     

Overall, HGs including HG 031203 
as well as the five low-metallicity candidate AGNs from \citet{IT08} and  
\citet{I10} lie in the region occupied by LCGs,
implying the similarity of their properties.
The location of HG 031203 in the upper left region of BPT diagram 
can be accounted for by the young age of the starburst. Starbursts with older 
age are instead located in the right lower part of the SFG region.

\subsection{Element abundances \label{abund}}


\begin{table*}
\caption{Emission-line fluxes \label{tab1}}
\begin{tabular}{lrrrr|lrrrr} \hline\hline
Line                 &$F$($\lambda$)&$I$($\lambda$)&Case B$^a$&Cloudy&
Line                 &$F$($\lambda$)&$I$($\lambda$)&Case B$^a$&Cloudy \\ \hline
\multicolumn{5}{c|}{a) Near-UV and optical range}                          &7281 He {\sc i}        &  1.6$\pm$0.4&  0.4$\pm$0.1&      &  0.6  \\
3726 [O {\sc ii}]    & 11.4$\pm$0.5& 40.2$\pm$1.7&       & 34.5&7320 [O {\sc ii}]      &  6.2$\pm$0.4&  1.5$\pm$0.1&      &  1.4  \\
3729 [O {\sc ii}]    & 12.9$\pm$0.5& 45.6$\pm$1.8&       & 40.4&7330 [O {\sc ii}]      &  6.1$\pm$0.4&  1.4$\pm$0.1&      &  1.1  \\
3750 H12             &  1.0$\pm$0.2&  6.5$\pm$1.1&  3.07 &  3.6&7751 [Ar {\sc iii}]    &  7.0$\pm$0.4&  1.3$\pm$0.1&      &  1.2  \\
3770 H11             &  1.0$\pm$0.3&  6.3$\pm$1.3&  4.00 &  4.5&8548 Pa15              &  5.4$\pm$0.5&  1.1$\pm$0.1&  0.53&  0.5  \\
3797 H10             &  1.5$\pm$0.3&  7.8$\pm$1.0&  5.34 &  5.8&8601 Pa14              &  6.2$\pm$0.4&  1.2$\pm$0.1&  0.65&  1.0  \\
3835 H9              &  2.4$\pm$0.5&  9.6$\pm$1.5&  7.37 &  7.8&8753 Pa12              &  6.1$\pm$0.7&  1.2$\pm$0.1&  1.04&  1.2  \\ 
3868 [Ne {\sc iii}]  &  7.1$\pm$0.7& 21.0$\pm$1.5&       & 60.6&8865 Pa11              &  5.7$\pm$0.4&  1.1$\pm$0.1&  1.35&  1.5  \\
3889 He {\sc i}+H8   &  4.9$\pm$0.4& 16.6$\pm$1.1& 10.60 & 17.2&9017 Pa10              &  9.2$\pm$1.2&  1.5$\pm$0.2&  1.80&  2.0  \\
3967 [Ne {\sc iii}]  &  6.3$\pm$0.4& 16.4$\pm$0.9&       & 18.3&9069 [S {\sc iii}]     &188.8$\pm$6.7& 24.8$\pm$0.7&      & 18.6  \\
3970 H7              &  4.0$\pm$0.2& 13.4$\pm$0.9& 16.00 & 16.5& \multicolumn{5}{c}{b) NIR range}                           \\
4026 He {\sc i}      &  1.5$\pm$0.4&  3.7$\pm$0.7&       &  1.8&9069 [S {\sc iii}]     &118.0$\pm$4.2& 15.0$\pm$0.4&      & 18.6  \\
4068 [S {\sc ii}]    &  1.3$\pm$0.4&  3.0$\pm$0.6&       &  0.9&9232 Pa9               & 20.7$\pm$1.8&  2.8$\pm$0.2&  2.49&  2.7  \\
4101 H$\delta$       &  9.8$\pm$0.6& 23.6$\pm$1.2& 26.10 & 26.5&9532 [S {\sc iii}]     &329.1$\pm$5.8& 37.6$\pm$0.6&      & 46.1  \\
4340 H$\gamma$       & 29.3$\pm$1.0& 50.0$\pm$1.4& 47.10 & 47.4&9549 Pa$\epsilon$      & 28.8$\pm$0.9&  3.7$\pm$0.1&  3.57&  3.7  \\
4363 [O {\sc iii}]   &  5.3$\pm$0.5&  8.5$\pm$0.6&       &  7.4&10052 Pa$\delta$       & 40.2$\pm$2.1&  4.4$\pm$0.2&  5.40&  5.6  \\
4388 He {\sc i}      &  0.9$\pm$0.4&  1.3$\pm$0.4&       &  0.5&10829 He {\sc i}       &295.1$\pm$5.2& 26.8$\pm$0.5&      & 31.4  \\
4471 He {\sc i}      &  2.8$\pm$0.3&  3.9$\pm$0.4&       &  3.8&10941 Pa$\gamma$       & 76.8$\pm$1.5&  7.0$\pm$0.2&  8.77&  9.0  \\
4711 [Ar {\sc iv}]   &  1.6$\pm$0.3&  1.8$\pm$0.2&       &  0.7&12821 Pa$\beta$$^b$    &125.5$\pm$2.7&  9.0$\pm$0.2& 15.70& 16.0  \\
4713 He {\sc i}      &  0.9$\pm$0.2&  1.1$\pm$0.2&       &  0.5&15884 Br14             &  4.8$\pm$0.3&  0.4$\pm$0.0&  0.31&  0.5  \\
4861 H$\beta$        &100.0$\pm$2.0&100.0$\pm$1.3&100.00 &100.0&16114 Br13             &  1.8$\pm$0.2&  0.2$\pm$0.0&  0.39&  0.5  \\
4959 [O {\sc iii}]   &283.7$\pm$5.5&259.0$\pm$3.4&       &220.8&16412 Br12             & 14.8$\pm$0.5&  0.8$\pm$0.0&  0.50&  0.6  \\
5007 [O {\sc iii}]   &817.1$\pm$13.&719.8$\pm$8.2&       &664.6&17006 He {\sc i}       &  6.3$\pm$0.4&  0.3$\pm$0.0&      &  0.3  \\
5016 He {\sc i}      &  2.4$\pm$0.4&  2.1$\pm$0.3&       &  2.3&18179 Br9$^b$          & 15.7$\pm$0.8&  0.8$\pm$0.0&  1.21&  1.3  \\
5876 He {\sc i}      & 22.6$\pm$0.6& 11.7$\pm$0.3&       &  9.7&18756 Pa$\alpha$       &617.0$\pm$11.& 31.0$\pm$0.6& 31.90& 32.9  \\
6300 [O {\sc i}]     &  6.3$\pm$0.5&  2.6$\pm$0.2&       &  1.8&19451 Br$\delta$       & 34.7$\pm$0.9&  1.8$\pm$0.1&  1.73&  1.8  \\
6312 [S {\sc iii}]   &  3.8$\pm$0.4&  1.5$\pm$0.1&       &  1.8& \multicolumn{5}{c}{c) MIR range}                                 \\
6363 [O {\sc i}]     &  2.0$\pm$0.3&  0.8$\pm$0.1&       &  0.6&10.51$\mu$m [S {\sc iv}]  &47.2$\pm$15.&47.2$\pm$15.&     & 84.2  \\
6548 [N {\sc ii}]    & 12.9$\pm$0.6&  4.6$\pm$0.2&       &  2.6&12.81$\mu$m [Ne {\sc ii}] & 3.9$\pm$1.9& 3.9$\pm$1.9&     &  1.1  \\
6563 H$\alpha$       &807.6$\pm$14.&284.8$\pm$3.7&282.00 &285.2&15.56$\mu$m [Ne {\sc iii}]&58.9$\pm$6.1&58.9$\pm$6.1&     & 51.2  \\
6583 [N {\sc ii}]    & 37.6$\pm$0.9& 13.1$\pm$0.3&       &  7.7&18.71$\mu$m [S {\sc iii}] &28.9$\pm$3.9&28.9$\pm$3.9&     & 20.7  \\
6678 He {\sc i}      & 11.0$\pm$0.5&  3.6$\pm$0.2&       &  2.6& \\
6717 [S {\sc ii}]    & 26.5$\pm$0.7&  8.5$\pm$0.2&       &  6.1&$C$(H$\beta$)          &\multicolumn{2}{c}{1.67$\pm$0.02}   \\
6731 [S {\sc ii}]    & 21.2$\pm$0.6&  7.1$\pm$0.2&       &  5.0&EW(H$\beta$)$^c$       &\multicolumn{2}{c}{134$\pm$3}       \\
7065 He {\sc i}      & 12.1$\pm$0.5&  3.2$\pm$0.1&       &  4.5&$F$(H$\beta$)$^d$      &\multicolumn{2}{c}{3.85$\pm$0.04}   \\
7136 [Ar {\sc iii}]  & 23.3$\pm$1.4&  6.0$\pm$0.3&       &  4.8&EW(abs)$^c$            &\multicolumn{2}{c}{2.0$\pm$0.4}     \\
\hline \\
\end{tabular}

$^a$ hydrogen recombination-line relative intensities by \citet{HS87}.\\
$^b$ affected by the telluric absorption.\\
$^c$ in \AA. \\
$^d$ in units 10$^{-16}$ erg s$^{-1}$ cm$^{-2}$. 
\end{table*}


We derived element abundances from emission-line fluxes 
using a classical semi-empirical method. These lines trace the ISM
of HG 031203. The fluxes in all spectra were 
measured using Gaussian fitting with the IRAF SPLOT routine. 
The line flux errors 
include statistical errors derived with SPLOT
from non-flux-calibrated spectra, in addition to errors introduced
by the absolute flux calibration, which we
set to 1\% of the line fluxes, according to the uncertainties of 
absolute fluxes of relatively bright standard stars 
\citep{Oke1990,Colina1994,Bohlin1996,IT04a}. 
These errors will be propagated below into the calculation of 
the electron temperatures, the electron number densities, and the ionic and 
total element abundances.
  Given a function $f$($x,y,...,z$), the uncertainty $\sigma$$(f)$
is calculated as

\begin{equation}
\sigma(f)=\sqrt{\left(\frac{df}{dx}\sigma(x)\right)^2 + 
\left(\frac{df}{dy}\sigma(y)\right)^2 + ...
+ \left(\frac{df}{dz}\sigma(z)\right)^2}.
\label{sigma}
\end{equation}

The fluxes were corrected for extinction, using the reddening curve
of \citet{C89}, and for underlying
hydrogen stellar absorption \citep{ITL94}. 
The equivalent width of the H$\beta$ emission line is EW(H$\beta$) =
134 \AA\ corresponding to an age of 3-4 Myr. At this age the predicted
equivalent width of the Balmer absorption lines is 
EW(abs) $\sim$ 2-3\AA\ \citep{GD1999,Guseva2003}
with relatively small variations between different hydrogen lines.
Therefore we assume that EW(abs) is the same
for all hydrogen lines in the same galaxy but varies from galaxy to galaxy. 
This assumption is justified by the evolutionary stellar population synthesis 
models of \citet{GD1999,GD05}. We also note that 20 - 30\% uncertainties
in the EW(abs) for the 3 - 4 Myr starburst result in $<$ 0.2\% 
uncertainties in the fluxes of
the strongest emission lines H$\beta$ and H$\alpha$, which are most important
for the determination of the extinction coefficient.
The extinction coefficient 
$C$(H$\beta$) and equivalent widths of the hydrogen absorption lines
EW(abs) are calculated simultaneously, which minimises the deviations 
of corrected fluxes $I(\lambda)$/$I$(H$\beta$) of all hydrogen Balmer lines
from their theoretical recombination values as
\begin{eqnarray*}
\frac{I(\lambda)}{I({\rm H}\beta)} & = &\frac{F(\lambda)}{F({\rm H}\beta)}
\frac{EW(\lambda)+EW(abs)}{EW(\lambda)}\frac{EW({\rm H}\beta)}{EW({\rm H}\beta)+EW(abs)} \\
                       & \times & 10^{C({\rm H}\beta)f(\lambda)}.
\end{eqnarray*} 
Here $f$($\lambda$) is the reddening function
normalised to the value at the wavelength of the H$\beta$ line,
$F$($\lambda$)/$F$(H$\beta$) are the 
observed  hydrogen Balmer emission line fluxes relative to that
of H$\beta$, EW($\lambda$) and EW(H$\beta$)
the equivalent widths of emission lines, and EW(abs)
the equivalent widths of hydrogen absorption lines.
The extinction-corrected fluxes of emission lines other than 
those of hydrogen are derived from the equation 
\begin{eqnarray*}
\frac{I(\lambda)}{I({\rm H}\beta)} & = &\frac{F(\lambda)}{F({\rm H}\beta)}
                       \times 10^{C({\rm H}\beta)f(\lambda)} .
\end{eqnarray*} 

  The derived $C$(H$\beta$)
is applied for correction of all emission-line fluxes in the
entire wavelength range $\lambda$$\lambda$3200 -- 24000\AA.

We varied $R_V$, the ratio of total to selective extinction,
in the range from 2.5 to 4.0 and found that 
these variations change the relative line fluxes by less than $\sim$2\%.
In the following we adopt $R_V$ = 3.2. The extinction-corrected relative 
fluxes 100$\times$$I$($\lambda$)/$I$(H$\beta$) of the lines, 
the extinction coefficient
$C$(H$\beta$), the equivalent width of the H$\beta$ emission line 
EW(H$\beta$), the H$\beta$ observed flux $F$(H$\beta$), and the 
equivalent width of the 
underlying hydrogen absorption lines EW(abs) are given in Table \ref{tab1}.
 We note that fluxes of Balmer hydrogen emission lines corrected for 
extinction and underlying hydrogen absorption (column 3 in Table \ref{tab1}) 
are close to the theoretical recombination values of \citet{HS87} 
(column 4 of the Table), suggesting 
that the extinction coefficient $C$(H$\beta$) is derived reliably.

The physical conditions and the ionic and total heavy element 
abundances in the ISM of the GRB 031203 host galaxy were derived 
following \citet{I06a} (Table \ref{tab2}). 
In particular, we adopt
the temperature $T_e^{dir}$(O~{\sc iii}) for  
O$^{2+}$, Ne$^{2+}$, and Ar$^{3+}$, which is directly derived from the 
[O~{\sc iii}] $\lambda$4363/($\lambda$4959 + $\lambda$5007)
emission-line ratio.
 The electron temperatures $T_e$(O~{\sc ii}) and
$T_e$(S {\sc iii}) were derived from the empirical relations by
\citet{I06a} based on a photoionisation model of H {\sc ii} regions.
We also derive $T_e^{dir}$(O~{\sc ii})=(5211$\pm$32) K with direct methods 
based on extinction-corrected fluxes of emission lines  
from the [O~{\sc ii}]$\lambda$(3726+3729)/$\lambda$(7320+7330) 
emission-line ratio,
$T_e^{dir}$(S~{\sc iii})=(9488$\pm$821) K from the 
[S~{\sc iii}]$\lambda$9069/$\lambda$6312 emission-line ratio
and $T_e^{dir}$(S~{\sc iii})=(13282$\pm$449) K from the
[S~{\sc iii}]$\lambda$(9069+9532)/$\lambda$6312 emission-line ratio.
We used $T_e$(O~{\sc ii}) for the calculation of
O$^{+}$,  N$^{+}$ and S$^{+}$ abundances and $T_e$(S {\sc iii})
for the calculation of S$^{2+}$ and Ar$^{2+}$ abundances.
 The electron number densities  $N_e^{dir}$(O~{\sc ii})=223$\pm$12 and 
$N_e^{dir}$(S~{\sc ii})=234$\pm$65 were obtained from the 
[O~{\sc ii}]$\lambda$3726/$\lambda$3729 and 
[S~{\sc ii}]$\lambda$6717/$\lambda$6731 emission-line ratios, respectively.
Therefore, the low-density limit holds for the H~{\sc ii} regions
that exhibit the narrow line components considered here. 
Then, the element abundances do not depend sensitively on $N_e$.
 The electron temperatures $T_e^{dir}$(O~{\sc iii}), 
$T_e$(O~{\sc ii}), $T_e$(S {\sc iii}), $T_e^{dir}$(S {\sc iii}) and 
$T_e^{dir}$(O~{\sc ii}), 
the electron number densities $N_e^{dir}$(O~{\sc ii}) and 
$N_e^{dir}$(S~{\sc ii}), the ionisation correction factors ($ICF$s), and
the ionic and total O, N, Ne, S and Ar abundances derived from the
forbidden emission lines are given in Table \ref{tab2}.

\begin{figure}
\hspace*{0.0cm}\psfig{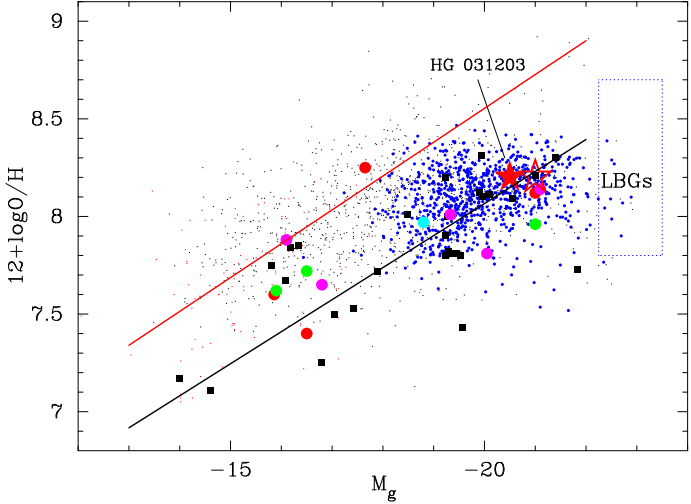}
\caption{Luminosity-metallicity relation 
for LCGs from 
SDSS DR7 \citep{IGT2010} (small blue circles).   
 The location of HG 031203
is shown by the large filled red star. 
 By filled squares are denoted the three 
most metal-poor BCDs (SBS 0335--052E, SBS 0335--052W and I Zw 18) 
\citep{G2009}, the intermediate-redshift ($z$ $<$ 1) extremely 
low-metallicity emission-line galaxies studied by \citet{Kakazu2007},
and luminous metal-poor star-forming galaxies at $z$ $\sim$ 0.7 studied
by \citet{Hoyos2005}. The position of the Lyman-break 
galaxies at $z$ $\sim$ 3 of \citet{Pettini2001} 
is marked as LBG with a dotted-line rectangle in the upper right.
The emission-line galaxies studied by \citet{G2009} are shown by
dots (red dots are data from our observations and 
black dots are SDSS data).
 The best linear-likelihood fit to the strongly star-forming galaxies
derived by \citet{IGT2010} is shown by the black straight line. 
The fit to emission-line galaxies studied by \citet{G2009} is shown by 
the red line.
Local ($z <$ 0.2) HGs 
compiled by \citet{Margutti2007} are shown by filled red circles.  
Data from \citet{Levesque2010}, \citet{Hammer2006} and \citet{Han2010} 
are shown by green, light blue and purple circles, respectively.
\vspace{0.1cm}
\hspace{10.5cm} (A colour version of this figure is available in the online journal.)
}
\label{LZ}
\end{figure}



\begin{table}[t]
\caption{Physical conditions and element abundances \label{tab2}}
\begin{tabular}{lc} \hline\hline
Property     &Value         \\ \hline 

$T_e^{dir}$(O {\sc iii})(4959+5007)/4363, K   &12161$\pm$307  \\
$T_e$(O {\sc ii}), K                   &12034$\pm$285  \\
$T_e$(S {\sc iii}), K                  &12098$\pm$255  \\
$T_e^{dir}$(S {\sc iii})9069/6312, K            &9488$\pm$821 \\
$T_e^{dir}$(S {\sc iii})(9069+9532)/6312, K    &13282$\pm$449 \\
$T_e^{dir}$(O {\sc ii})(3726+3729)/(7320+7330)), K  &5211$\pm$32 \\ \\
$N_e^{dir}$(O {\sc ii})3726/3729, cm$^{-3}$      &223$\pm$12    \\
$N_e^{dir}$(S {\sc ii})6716/6731, cm$^{-3}$      &234$\pm$65   \\ \\
O$^+$/H$^+$, ($\times$10$^5$)          &1.64$\pm$0.14  \\
O$^{2+}$/H$^+$, ($\times$10$^5$)       &14.20$\pm$1.06  \\
O/H, ($\times$10$^5$)                  &15.84$\pm$1.07  \\
12+log O/H                             &8.20$\pm$0.03  \\ \\
N$^{+}$/H$^+$, ($\times$10$^6$)        &1.55$\pm$0.08  \\
$ICF$(N)$^a$                           &     7.82      \\
N/H, ($\times$10$^6$)                  &12.09$\pm$0.78  \\
log N/O                                &--1.12$\pm$0.04~~\\ \\
Ne$^{2+}$/H$^+$, ($\times$10$^5$)      &3.15$\pm$0.28  \\
$ICF$(Ne)$^a$                          &      1.04     \\
Ne/H, ($\times$10$^5$)                 &3.30$\pm$0.31  \\
log Ne/O                               &--0.68$\pm$0.05~~\\ \\
S$^{+}$/H$^+$, ($\times$10$^6$)        &0.24$\pm$0.01  \\
S$^{2+}$/H$^+$, ($\times$10$^6$)       &1.64$\pm$0.17  \\
$ICF$(S)$^a$                           &      2.09     \\
S/H, ($\times$10$^6$)                  &3.93$\pm$0.36  \\
log S/O                                &--1.60$\pm$0.05~~\\ \\
Ar$^{2+}$/H$^+$, ($\times$10$^7$)      &4.49$\pm$0.20  \\
$ICF$(Ar)$^a$                          &      1.24     \\
Ar/H, ($\times$10$^7$)                 &5.58$\pm$0.23  \\
log Ar/O                               &--2.57$\pm$0.04~~\\ \\
\hline
\end{tabular}

\smallskip

$^a$ Ionisation correction factor.
\end{table}


Our oxygen abundance of HG 031203
12+logO/H = 8.20$\pm$0.03 generally agrees
with previous direct determinations of 12 + log O/H in the range 
$\sim$ 8.0 -- 8.2
\citep{Prochaska2004,Hammer2006,Margutti2007,Wiersema2007,Levesque2010,Han2010}
(Fig.~\ref{oiioiii}). In particular,
 the oxygen abundance 12+logO/H = 8.20$\pm$0.03 is consistent
with 8.19 estimated by \citet{Sollerman2005} from the thorough analysis 
presented by \citet{Prochaska2004}.
On the other hand, our value is slightly higher than the 
12 + log O/H = 8.12 $\pm$ 0.04 obtained by 
\citet{Margutti2007} (VLT optical spectra),
8.14 $\pm$0.07 by \citet{Han2010}
and  higher than the 12 + log O/H = 8.02 $\pm$ 0.15 derived by
\citet{Prochaska2004} (Magellan/IMACS optical spectra) and 
7.96 derived by \citet{Levesque2010}.

We also recalculated 12 + log O/H with emission-line
fluxes given by \citet{Margutti2007}, \citet{Han2010}, 
\citet{Prochaska2004} and \citet{Levesque2010}
and obtained 8.22$\pm$0.03, 8.15$\pm$0.01, 8.06$\pm$0.08 and 8.02, 
respectively, which are higher by $\sim$ 0.01 - 0.10 dex that the values
in the original papers.

  Abundance ratios log N/O, log Ne/O, log S/O and
log Ar/O vs. oxygen abundance 12 + log O/H  
for HG 031203 (this paper) and for LCGs from SDSS DR7 \citep{IGT2010}  
are shown in Fig.~\ref{oiioiii}
by large red filled stars and  small blue filled circles,
respectively. We also plot the available abundance ratios from the literature 
and recalculated values from published emission-line fluxes  
by filled and open symbols, respectively.
Data for HG 031203 are shown by stars of different colours. 
Specifically, we label data of 
\citet{Margutti2007} by filled and open purple stars (HG 031203), 
\citet{Hammer2006} by filled and open black circles 
(HG 980425, SN and WR regions, HG 020903), 
\citet{Wiersema2007} by filled yellow triangles (HG 060218), 
\citet{Levesque2010} by open light blue squares 
(HGs 031203, 030329 and 060218), 
\citet{Han2010} by open blue rhombs (HGs 031203, 030329, 020903 and 
060218).
  Data for HG 031203 from \citet{Prochaska2004} 
(using solar relative abundances 
by \citet{Grevesse1996} or by \citet{Holweger2001}) are shown by  
filled green large and small stars, respectively.
 Abundance ratios obtained by \citet{Prochaska2004},
\citet{Margutti2007}, \citet{Hammer2006} 
and our redetermined data obtained from their emission line fluxes 
are connected by straight lines.
 The solar abundance ratios by \citet{Asplund2009} are indicated by the large 
purple open circles and the associated error bars. 
We use only those data where $T_e$-sensitive 
[O {\sc iii}]$\lambda$4363\AA\ emission line is measured.

\citet{Wiersema2007} found the enhanced N/O ratio for several HGs
as compared to that in emission-line galaxies.
However, our value of log N/O for HG 031203 (Fig.~\ref{oiioiii}a) is 
lower by 0.3 - 0.4 dex than log N/O = -0.74$\pm$0.20 by
\citet{Wiersema2007} and log N/O = -0.83$\pm$0.20 by \citet{Prochaska2004},
but it is by 0.4 dex higher than log N/O = -1.54$\pm$0.08 by 
\citet{Margutti2007}.
 For HG 060218 \citet{Wiersema2007} obtained log N/O = -1.0$\pm$0.4 
(our recalculation gives log N/O = -1.19$\pm$0.88).    
  As Fig.~\ref{oiioiii}a shows the larger sample of HGs
overlaps
emission-line galaxies from SDSS DR7. For HG 031203 and HG 060218 and 
for some  emission-line galaxies from SDSS DR7 N/O ratio is higher. 
This may be owing to local 
enrichment of the ISM by GRB progenitors or SN remnants.

From the observed spectrum of HG 031203 we derived 
log Ne/O = --1.15$\pm$0.06, 
using the [Ne {\sc iii}]$\lambda$3868\AA\ emission line, which is too low 
compared to the log Ne/O abundance ratio in other emission-line galaxies.
This is because the flux ratio  
[Ne {\sc iii}]$\lambda$3868/$\lambda$3967 of 1.13 is very low 
because the [Ne {\sc iii}]$\lambda$3868\AA\ line is located in a very noisy
part of the spectrum which renders its flux uncertain. 
For comparison, the CLOUDY-predicted line ratio for 
[Ne {\sc iii}]$I$($\lambda$3868)/$I$($\lambda$3967) is 
$\sim$3.32 (or $F$($\lambda$3868)/$F$($\lambda$3967) $\sim$2.93, if 
$C$(H$\beta$)=1.67 and the reddening function of \citet{C89} are used). 
 Therefore, we use 
$F(3868)$ = $F(3967)$ $\times$ 2.93 for the Ne abundance determination.
Then log Ne/O = --0.68$\pm$0.05 (see Table~\ref{tab2}), which is 
consistent with the data for LCGs \citep{IGT2010}.

\citet{Margutti2007} do not describe how they derived the 
sulphur abundance. We suspect that their very low sulphur abundance 
shown in Fig.~\ref{oiioiii}c may have been derived because the
[S {\sc iii}]$\lambda$6312\AA\ emission line was not included for the sulphur
abundance determination. To demonstrate this suspicion we show by smaller open
symbols in Fig.~\ref{oiioiii}c a sulphur abundance without the
[S {\sc iii}]$\lambda$6312\AA\ emission line. 
These determinations result in low log S/O $\la$ --2.0 and are out of 
the general relation.
  Note that our recalculated S/O abundance ratios (in galaxies with 
detected [S {\sc iii}]$\lambda$6312\AA\ line
(\citet{Margutti2007} and \citet{Prochaska2004} for HG 031203 and 
\citet{Hammer2006} for HG 980425) 
are consistent with the S/O determinations for other emission-line galaxies.

Overall, element abundance ratios for a sample of HGs and our 
log N/O = --1.12$\pm$0.04, log Ne/O = --0.68$\pm$0.05, 
log S/O = --1.60$\pm$0.05
and log Ar/O = --2.57$\pm$0.04 for HG 031203 agree well
with the abundance ratios obtained for a large sample of 
803 spectra of LCGs \citep{IGT2010}.

\subsection{Extinction and hidden star formation\label{extinc}}

That the spectrum of HG 031203 has been obtained 
simultaneously over the entire optical and near-infrared wavelength  
ranges eliminates uncertainties introduced by the use of different apertures 
in the separate optical and NIR observations. 
In all previous studies of low-metallicity emission-line galaxies except for 
that of Mrk 59 in 
\citet{I09}, II Zw 40, Mrk 71, Mrk 996, SBS 0335--052E in
\citet{IT11}  and PHL 293B in \citet{I10a} the NIR spectra
were obtained in separate $JHK$
observations, and there was no wavelength overlap between the 
optical and NIR spectra. The elimination of these adjusting uncertainties 
permits us to compare directly the
extinctions derived from the optical and NIR spectra.
 
In the second column of Table \ref{tab1} we show the observed 
fluxes $F$($\lambda$) of the emission lines. The extinction coefficient 
$C$(H$\beta$) and the equivalent width
EW(abs) of hydrogen lines are derived from the decrement of hydrogen Balmer
lines in the UVB and VIS spectra. 
  The extinction coefficient $C$(H$\beta$) = 1.67
is adopted for correction of emission line fluxes in 
all UVB, VIS, and NIR ranges. In the third column
of the same table the corrected fluxes $I$($\lambda$) are shown.
  In the fourth column of the table the theoretical recombination fluxes 
for hydrogen lines calculated by \citet{HS87} are shown.
 The relative intensities of hydrogen recombination lines \citep{HS87} 
are calculated for an electron temperature $T_e$ = 12500 K, an electron
number density $N_e$ = 100 cm$^{-3}$, and the case B theory.

  Note that the comparison of the observed fluxes of Balmer, 
Paschen, and Brackett hydrogen lines after correction for  
extinction with a single value $C$(H$\beta$) = 1.67 
and the equivalent width EW(abs) of hydrogen lines of 2.0\AA\ 
and the theoretical recombination values \citep{HS87} 
shows an agreement within 2 - 5\% for strongest H$\delta$, H$\gamma$,
H$\alpha$, and Pa$\alpha$ lines.
On the other hand, the differences between extinction-corrected and predicted
intensities for some lines are high (e.g., for Pa$\beta$ line) because of the
strong telluric absorption.
The good agreement implies that a single $C$(H$\beta$) can be used over the 
whole $\sim$$\lambda\lambda$3200 -- 24000\AA\ range to correct line 
fluxes for extinction. 
  That the extinction coefficient $C$(H$\beta$)
does not increase when moving from the optical to the 
NIR wavelength ranges implies that the NIR emission lines arise in the regions 
where extinction is not as high and that the NIR emission does not have
a hidden star formation compared to the optical emission 
lines. This appears to be a general result for low-metallicity emission-line
galaxies \citep[e.g.][]{V00,V02,I09,I10a,IT11}.

To estimate the intrinsic host-galaxy extinction we used the far-IR map 
by \citet{Schlegel1998}, from which we inferred the reddening value for the 
Milky Way, $E_{\rm MW}$($B-V$) $\approx$ 1.037 [or $C$(H$\beta$)$_{\rm MW}$ 
= 1.52].
  We derived the estimate of the intrinsic reddening through comparison of 
$C$(H$\beta$)$_{\rm HG}$=1.67 derived from Balmer decrement in the spectrum 
and $C$(H$\beta$)$_{\rm MW}$ = 1.52 by \citet{Schlegel1998}. 
Thus, the intrinsic $C$(H$\beta$)$_{\rm HG}$ = 0.15 [or $A(V)$$_{(\rm HG)}$ = 0.33].
This is typical for emission-line dwarf galaxies, but lower than 
$C$(H$\beta$)$_{\rm HG}$ = 0.59 by \citet{Margutti2007} and 
$C$(H$\beta$)$_{\rm HG}$ = 0.46 deduced by \citet{Prochaska2004}.
Recently \citet{Han2010} obtained $C$(H$\beta$)$_{\rm HG}$=0.13 $\pm$ 0.01.
\citet{Levesque2010} derived the total colour excess in the direction 
of the galaxy $C$(H$\beta$) = 1.72. If the $C$(H$\beta$)$_{\rm MW}$ 
= 1.52 from \citet{Schlegel1998} is adopted, $C$(H$\beta$)$_{\rm HG}$ is
0.2. Both these determinations are close to our value of intrinsic extinction.

\subsection{CLOUDY stellar photoionisation modelling of the H {\sc ii} region
and mid-infrared emission lines \label{CLOUDY}}

We next examine the excitation mechanisms of all emission 
lines (not only hydrogen lines) arising in the H {\sc ii} region of HG 031203. 
For this purpose, we have constructed 
stellar photoionisation model for the H {\sc ii} region using
the CLOUDY code (version 07.02.01) of \citet{F98}.  
We list in Table \ref{tab5} the input parameters of this model. 
The first row gives $Q$(H), the logarithm of
the number of ionising photons per second, 
calculated from the extinction-corrected H$\beta$ luminosity.
The other rows of Table \ref{tab5} list starburst age with the spectral
energy distribution of the ionising stellar
radiation calculated with Starburst-99 models \citep{L99}, and $N_e$, 
the electron number density, assumed to be
constant with radius. The fourth row gives $f$, the filling factor. 
The remaining input parameters are the ratios of number densities of
different species to hydrogen, derived 
from the spectroscopic observations in this paper, except for the carbon
abundance. There are no strong emission lines of carbon 
in the optical and NIR ranges. We therefore calculated 
its abundance using the mean relation of C/O vs. oxygen abundance derived 
for low-metallicity H {\sc ii} regions by \citet{G95a}.
Parameters $Q$(H), $N_e$ and chemical composition are derived from 
the spectra and thus are kept unchanged. The starburst age and filling factor
are not constrained by the observational data. 
Therefore, we varied these two parameters to achieve the
best agreement between observations and model predictions. Their adopted values
are shown in Table \ref{tab5}.

The CLOUDY-predicted fluxes of the  
emission lines are shown in the fifth column of Table \ref{tab1}. 
Comparison of the extinction-corrected observed
and predicted emission-line fluxes
in both the optical and NIR ranges shows that in general, the agreement 
between the observations and the CLOUDY predictions is
achieved.
This implies that a H {\sc ii} region model including only stellar 
photoionisation as ionising source is able to 
account for the observed fluxes both in the optical and NIR ranges.
No additional excitation mechanism such as shocks from stellar winds and 
supernova remnants is needed.

We showed above that the observed emission-line fluxes 
in the optical and NIR ranges in HG 031203 can be 
satisfactorily accounted for with the 
same extinction coefficient $C$(H$\beta$) [or the same extinction
$A$($V$)]. In other words,  
there is no more hidden star formation in the NIR range as 
in the optical range in regions traced
by emission lines, i.e. in ionised gas regions.
A question then arises: would we see more hidden star formation
at longer wavelengths, in the mid-infrared range?  
To investigate this question in the MIR range,  
we used the {\sl Spitzer} observations
of \citet{Watson2010}. The MIR emission line fluxes relative to 
the H$\beta$ emission line flux and CLOUDY predicted fluxes are shown
in Table \ref{tab1}. No extinction correction was applied to the 
observed MIR data because it is presumably small in this wavelength range.
Comparison of the MIR observed fluxes
with those predicted by the pure stellar ionising radiation model 
shows agreement to within a factor of $\sim$2 or better. 
The disagreement is the worst for the very weak low-ionisation line 
[Ne {\sc ii}] $\lambda$12.81 $\mu$m, whose flux is predicted 
to be considerably lower than the observed one.

The agreement
between the observed fluxes of 
the MIR emission-lines and the predicted ones of CLOUDY model, 
based on the extinction-corrected optical emission lines 
with $C$(H$\beta$) = 1.67,
implies that there is only little more hidden star 
formation seen in the 
MIR range compared to that in the optical and NIR ranges.
This implies that in HG 031203 the MIR
emission lines  emerge in relatively transparent regions, which are also 
seen in the optical and NIR ranges. The same conclusion for five other
emission-line galaxies were made by \citet{I09} and \citet{IT11}.

\subsection{Kinematic structure \label{kinem}}

The brightest emission lines in the spectrum of HG 031203
significantly deviate from a single 
Gaussian line profile (Fig.~\ref{profiles}, right panel). We rule out 
instrumental effects as the cause because no deviations from single Gaussian 
profiles were detected in night-sky line profiles.
To study the kinematic structure of HG 031203 we 
reassembled all strong emission 
lines, obtaining  narrow and broad components separated 
by $\sim$34 km s$^{-1}$.
  Decomposition of H$\beta$, [O {\sc iii}] $\lambda$4959, $\lambda$5007, 
H$\alpha$, [S {\sc iii}] $\lambda$9532 and Pa$\alpha$ $\lambda$18756
emission line profiles into two Gaussian 
components is shown in Fig.~\ref{profiles}, left panel.
 The observed spectrum and the fit are shown by black solid and red 
dashed lines, respectively. Two Gaussian components and residual spectra
are shown by blue dashed and black dotted lines, respectively. 
For convenience the observed 
spectra, Gaussians, and residuals are shifted along the ordinate axis.
Dispersions $\sigma_{res}$ of residual spectra for two cases: 1) fit of 
observed 
profile by two Gaussians and 2) fit by single Gaussian, are presented in 
the Table~\ref{tab6}. This table and 
Fig.~\ref{profiles} show that the observed bright emission
lines are fitted much better by the two Gaussians.

\begin{table} [t]
\caption{Input parameters for stellar photoionisation 
CLOUDY model
 \label{tab5}}
\begin{tabular}{lc} \hline \hline
Parameter              &Value \\ \hline 
log $Q$(H)$^a$         & 53.73 \\
Starburst age, Myr     & 1.3 \\
$N_e$, cm$^{-3}$       & 250    \\
$f$$^b$                & 0.01   \\
log He/H               & --1.09 \\
log C/H                & --4.30 \\
log N/H                & --4.92 \\
log O/H                & --3.80 \\
log Ne/H               & --4.48 \\
log S/H                & --5.40 \\
log Ar/H               & --6.37 \\ \hline
\end{tabular}


$^a$$Q$(H) is the number of ionising photons per second. \\
$^b$Filling factor. \\
\end{table}

\begin{table}[t]
\caption{Dispersions of residual spectra\label{tab6}}
\begin{tabular}{lcrr} \hline\hline
  &  & \multicolumn{2}{c}{$\sigma$$_{res}$$^b$}\\ \cline{3-4}  
Line                    &\multicolumn{1}{c}{$\Delta$$\lambda$$^a$}&\multicolumn{1}{c}{double profile}
                        &\multicolumn{1}{c}{single profile}  \\ \hline 
4861 H$\beta$           & 4852-4870   & 0.015 & 0.034 \\
4959 [O {\sc iii}]      & 4951-4965   & 0.039 & 0.097 \\
5007 [O {\sc iii}]      & 4999-5013   & 0.084 & 0.230 \\
6563 H$\alpha$          & 6553-6572   & 0.076 & 0.296 \\
9532 [S {\sc iii}]      & 9520-9547   & 0.062 & 0.140 \\
18756 Pa$\alpha$        & 18730-18780 & 0.099 & 0.151 \\
\hline
\multicolumn{4}{l}{$^a$Wavelength range for the $\sigma$$_{res}$ determination.} \\
\multicolumn{4}{l}{$^b$Dispersions of residual spectra shown in 
Fig.~\ref{profiles}.} \\
\end{tabular}
\end{table}

In Table~\ref{tab3} we present observed fluxes, 
FWHM (in km s$^{-1}$)
and $F(\lambda$)/$F(H\beta$) ratios of the narrow and broad 
components for 14 strong emission lines. 
The velocity errors are obtained from statistical errors of FWHMs 
from non-flux-calibrated spectra using the IRAF SPLOT routine. 

The FWHMs of the narrow-line components and broad-line components in the 
spectrum of HG 031203 are $\sim$90--130 km s$^{-1}$ and 
$\sim$200--330 km s$^{-1}$, respectively.
The respective average FWHM values for narrow and broad components 
are $\sim$115 and $\sim$270 km s$^{-1}$. 
Note that the Pa$\alpha$ profile 
is likely modified by fringes showing a wavy structure 
(Fig.~\ref{profiles}f).
Therefore, the decomposition of the Pa$\alpha$ line
is not as accurate as that for other lines.
The two velocity components of strong emission lines
likely arose in two star-forming regions of the host galaxy. 
HG 031203 is a compact galaxy with a major axis of $\sim$1\arcsec, 
which is comparable to the slit width $\sim$1\arcsec. Therefore, it is 
complicated from its kinematical structure to distinguish between regular 
rotating disc motion and stochastic motion of H {\sc ii} regions 
\citep[e.g. ][]{Christensen2008,Thoene2008}.
A similar kinematic structure was found by \citet{Wiersema2007}
from two strong emission lines [O {\sc iii}] $\lambda$4959, $\lambda$5007 
in HG 060218. The authors detected two components separated by 
$\sim$ 22 km s$^{-1}$ that were 
repeated in Na {\sc i} and Ca {\sc ii} absorption lines.
The image of AGN candidate Tol 2240--384 \citep{I10} also reveals the 
two H {\sc ii} regions, 
separated by $\sim$80 km s$^{-1}$, if the spectral profiles of strongest 
emission lines are fitted by two Gaussians.

{\begin{table*}[t]
\caption{Observed fluxes of narrow and broad components of strong emission lines\label{tab3}}
\begin{tabular}{lrrlcrrlr} \hline\hline
    & \multicolumn{3}{c}{Narrow}&&\multicolumn{3}{c}{Broad}&\multicolumn{1}{c}{Whole line}\\ \cline{2-4} \cline{6-8} 
Line                    &\multicolumn{1}{c}{$F$($\lambda$)$^a$}&\multicolumn{1}{c}{FWHM$^b$}
                        &\multicolumn{1}{c}{$F$($\lambda$)/$F$(H$\beta$)}
                        &&\multicolumn{1}{c}{$F$($\lambda$)$^a$}&\multicolumn{1}{c}{FWHM$^b$}
                        &\multicolumn{1}{c}{$F$($\lambda$)/$F$(H$\beta$)}
                        &\multicolumn{1}{c}{$F$($\lambda$)/$F$(H$\beta$)}  \\ \hline 
4340 H$\gamma$          &  0.802$\pm$0.090 & 132$\pm$10&0.38$\pm$0.04&& 0.471$\pm$0.074& 330$\pm$42&   0.29$\pm$0.05&0.29$\pm$0.01 \\
4861 H$\beta$           &  2.100$\pm$0.042 & 131$\pm$3 &1.00$\pm$0.03&& 1.600$\pm$0.027& 290$\pm$4&     1.00$\pm$0.03&1.00$\pm$0.02 \\
4959 [O {\sc iii}]      &  6.330$\pm$0.121 & 131$\pm$2 &3.01$\pm$0.09&& 4.270$\pm$0.074& 290$\pm$4&     2.67$\pm$0.07&2.81$\pm$0.05 \\
5007 [O {\sc iii}]      & 14.700$\pm$0.136 & 127$\pm$1 &7.00$\pm$0.16&&12.000$\pm$0.117& 279$\pm$2&     7.50$\pm$0.18&8.08$\pm$0.14 \\
5876 He {\sc i}         &  0.493$\pm$0.041 & 126$\pm$7 &0.23$\pm$0.02&& 0.382$\pm$0.028& 315$\pm$18&    0.24$\pm$0.02&0.22$\pm$0.01 \\
6563 H$\alpha$          & 13.100$\pm$0.096 & 106$\pm$1 &6.24$\pm$0.16&&14.400$\pm$0.077& 280$\pm$1&     8.57$\pm$0.20&7.99$\pm$0.14 \\
6678 He {\sc i}         &  0.151$\pm$0.024 & 111$\pm$14&0.07$\pm$0.01&& 0.256$\pm$0.027& 290$\pm$30&   0.16$\pm$0.02&0.11$\pm$0.00 \\
6717 S {\sc ii}         &  0.525$\pm$0.028 & 103$\pm$4 &0.25$\pm$0.01&& 0.505$\pm$0.028& 261$\pm$10&    0.32$\pm$0.02&0.26$\pm$0.01 \\
6731 S {\sc ii}         &  0.358$\pm$0.020 &  95$\pm$5 &0.17$\pm$0.01&& 0.481$\pm$0.023& 257$\pm$11&    0.30$\pm$0.02&0.22$\pm$0.01 \\
7065 He {\sc i}         &  0.288$\pm$0.021 & 117$\pm$7 &0.14$\pm$0.01&& 0.149$\pm$0.028& 190$\pm$38&    0.09$\pm$0.02&0.12$\pm$0.00 \\
7136 [Ar {\sc iii}]     &  0.454$\pm$0.035 & 107$\pm$6 &0.22$\pm$0.02&& 0.452$\pm$0.041& 244$\pm$16&    0.28$\pm$0.03&0.23$\pm$0.01 \\
9532 [S {\sc iii}]      &  6.153$\pm$0.094 & 117$\pm$1 &2.93$\pm$0.08&& 6.304$\pm$0.097& 273$\pm$2&     3.94$\pm$0.09&3.25$\pm$0.06 \\
10829 He {\sc i}        &  3.045$\pm$0.050 &  91$\pm$1 &1.45$\pm$0.05&& 7.504$\pm$0.093& 221$\pm$1&     4.69$\pm$0.11&2.92$\pm$0.05 \\
18756 Pa$\alpha$        & 12.936$\pm$0.067 & 124$\pm$1 &6.16$\pm$0.15&& 9.872$\pm$0.073& 259$\pm$1&     6.17$\pm$0.14&6.10$\pm$0.11 \\
\hline
\multicolumn{9}{l}{$^a$Observed flux in units 10$^{-16}$ erg s$^{-1}$ cm$^{-2}$.} \\
\multicolumn{9}{l}{$^b$In km s$^{-1}$.} \\
\end{tabular}
\end{table*}
}


\begin{table*}
\caption{Parameters of narrow and broad components of strong Balmer lines\label{tab4}}
\begin{tabular}{lrr|rr|rr} \hline\hline
    & \multicolumn{2}{c}{Narrow}&\multicolumn{2}{c}{Broad}&\multicolumn{2}{c}{Whole}\\ \cline{2-3} \cline{4-5} \cline{6-7}
Line                 &$F$($\lambda$)/$F$(H$\beta$)&$I$($\lambda$)/$I$(H$\beta$)&
$F$($\lambda$)/$F$(H$\beta$)&$I$($\lambda$)/$I$(H$\beta$)& 
$F$($\lambda$)/$F$(H$\beta$)&$I$($\lambda$)/$I$(H$\beta$)\\ \hline
4340 H$\gamma$       & 0.38$\pm$0.04 & 0.56$\pm$0.05 &0.29$\pm$0.05& 0.50$\pm$0.06&0.29$\pm$0.01& 0.49$\pm$0.01  \\
4861 H$\beta$        & 1.00$\pm$0.04 & 1.00$\pm$0.05 &1.00$\pm$0.03& 1.00$\pm$0.06&1.00$\pm$0.02& 1.00$\pm$0.01  \\
6563 H$\alpha$       & 6.24$\pm$0.16 & 2.86$\pm$0.07 &8.57$\pm$0.20& 2.86$\pm$0.05&7.99$\pm$0.14& 2.85$\pm$0.04  \\
$C$(H$\beta$)&\multicolumn{2}{c}{1.25$\pm$0.03}    &\multicolumn{2}{c}{1.75$\pm$0.03} &\multicolumn{2}{c}{1.67$\pm$0.02} \\
EW(H$\beta$)$^a$&\multicolumn{2}{c}{70$\pm$1}      &\multicolumn{2}{c}{56$\pm$1}      &\multicolumn{2}{c}{134$\pm$3} \\
$F$(H$\beta$)$^b$&\multicolumn{2}{c}{2.10$\pm$0.04} &\multicolumn{2}{c}{1.68$\pm$0.03} &\multicolumn{2}{c}{3.85$\pm$0.04} \\
EW(abs)$^a$&\multicolumn{2}{c}{2.0$\pm$1.10}       &\multicolumn{2}{c}{2.0$\pm$0.73} &\multicolumn{2}{c}{2.0$\pm$0.4}  \\                       
\hline \\
\end{tabular}
\smallskip

$^a$ in \AA. \\
$^b$ in units 10$^{-16}$ erg s$^{-1}$ cm$^{-2}$. 

\end{table*}

\begin{figure}[t]
\hspace*{0.0cm}\psfig{figure=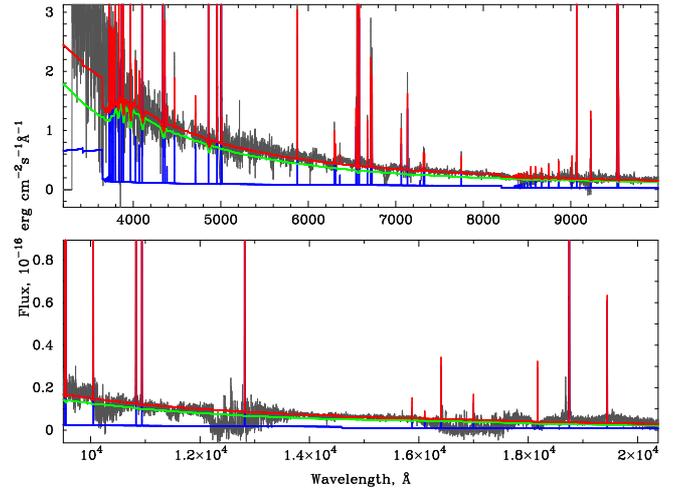,angle=-90,width=8.5cm,clip=}
\caption{Best-fit model SED to the redshift- and 
extinction-corrected observed spectrum. 
The contributions from the stellar and 
ionised gas components are shown by green and blue
lines, respectively. The sum of both stellar and ionised gas emission
is shown by the red line. Evidently,
 the spectrum in the whole wavelength range
$\sim$$\lambda$$\lambda$3200--24000\AA\ is fitted quite well despite
the strong absorption features in the NIR caused by 
telluric lines.
\vspace{0.1cm}
\hspace{12.5cm} (A colour version of this figure is available in the online journal.)
}
\label{sed}
\end{figure}

In Fig.~\ref{prof} the velocity excesses in the emission line profiles of 
H$\beta$ $\lambda$4861, [O {\sc iii}] $\lambda$4959, 
H$\alpha$ $\lambda$ 6563 and [O {\sc iii}] $\lambda$5007 are shown.
Observed profiles are denoted by solid lines and 
fits by dotted lines.
Fits of the narrow and broad components are shown in blue and summed 
fits in red. The positions of the blue and red velocity excess, corresponding 
to velocities of $v$=--400 km s$^{-1}$
and $v$=+350 km s$^{-1}$, respectively, are shown by vertical 
tick marks.
The bright Pa$\alpha$ line is not included in our analysis of the velocity 
excesses because it could be affected by fringes.
  This effect is clearly seen in the Pa$\alpha$ profile 
in Fig.~\ref{profiles}.
 Some traces of night-sky absorption lines at the position
of H$\alpha$ (the observed wavelength is 7255\AA) are seen 
in the blue part of the spectrum (Fig.~\ref{prof}).

  We cannot obtain physical conditions and heavy element 
abundances for the narrow and broad components separately
owing to the weakness of the key line [O {\sc iii}]$\lambda$4363. 
This precludes an accurate separation of narrow and 
broad components of this line. There is also insufficient spectral 
resolution for the decomposition of the 
[O {\sc ii}]$\lambda$3726+3729 emission lines 
into two pairs of narrow and broad components.
An inspection of Table~\ref{tab3} shows that the observed flux 
ratios $F$($\lambda$)/$F$(H$\beta$) are similar for the broad components and
the entire lines but differ from those of the narrow components.
    Therefore, we calculated the extinction coefficient $C$(H$\beta$)
of H$\gamma$, H$\beta$, and H$\alpha$ for the 
narrow-component region and obtained $C$(H$\beta$) = 1.24 (Table~\ref{tab4}).
  Assuming the same extinction for the region with the broad emission,
we obtain a broad H$\alpha$/H$\beta$ flux ratio of $\sim$ 4
or higher than the recombination ratio
of $\sim$ 2.8, suggesting either a higher density of the broad emission
region or a higher extinction compared to that in the
narrow-line region, or $R_V$ different from 3.2. Assuming the higher extinction
$C$(H$\beta$) = 1.75 for the broad-component region, we obtain robust estimates
of the Balmer-line fluxes (Table~\ref{tab4}).

\subsection{Luminosity-metallicity relation\label{LZr}}

Since the study by \citet{Lequeux1979} it was established that lower 
luminosity and lower stellar
mass galaxies also  have lower metallicities. This conclusion has
been confirmed using large samples of galaxies such as those from the SDSS 
\citep[e.g. ][]{Tremonti2004,G2009}.
Nevertheless, some local low-metallicity star-forming galaxies with strong
bursts of star formation (e.g. SBS 0335--052E,  
SBS 0335--052W) and some high-redshift galaxies, such as LBGs,
deviate from the luminosity-metallicity (L--Z) relation for the bulk
of emission-line galaxies, which are shifted to lower metallicities and/or to 
higher luminosities \citep{Kunth2000,G2009,IGT2010}.

Adopting  $B$=22.32 mag from VLT+FORS2 observation of HG 031203 
by \citet{Margutti2007} and $C$(H$\beta$) = 1.67 we 
obtain $B_{corr}$=17.66 mag. At a distance of 430 Mpc 
taken from the NASA/IPAC Extragalactic Database 
(NED)\footnote{NASA/IPAC Extragalactic Database (NED)
is operated by the Jet Propulsion Laboratory, California Institute of 
Technology, under contract with the National Aeronautics and Space 
Administration.} we derive an absolute magnitude $M_B$=--20.50. 
 The adopted distance was obtained from the radial velocity
corrected for Virgo infall with a Hubble constant 
of 73 km s$^{-1}$ Mpc$^{-1}$.
  For comparison, the absolute blue magnitude $M_B$ of HG 031203 from 
\citet{Kewley2007} is --19.3.  \citet{Savaglio2009}, who
estimated $M_B$ in the AB system, found --21.11.
  We compared the absolute magnitudes $M_B$ for 
HG 031203 with the SDSS $g$ absolute magnitudes $M_g$ for LCGs
from \citet{IGT2010}. This is possible thanks to the prescriptions 
of \citet{Papaderos2008}, who investigated the $B-g$ index and concluded
that it is very small and varying in the range of $\sim$0.01--0.03 mag  
for different star-formation histories of a galaxy.

Fig.~\ref{LZ} shows the relation between the oxygen abundance 
12 + log O/H and the
absolute magnitude $M_g$ in the SDSS $g$ band for the extensive
sample of emission-line galaxies studied by \citet{G2009} (red dots for 
our observations and black dots for SDSS sample) and for
LCGs from \citet{IGT2010} 
(small blue circles). The location of HG 031203
with absolute magnitude and oxygen abundance from this paper
is denoted by a large red star.
 The $g$ apparent magnitudes for LCGs and for emission-line
galaxies studied by \citet{G2009}
are taken from the SDSS.
The correction for extinction was made using extinction coefficients 
$C$(H$\beta$) of LCGs  derived from Balmer decrement in the SDSS spectra. 
Distances of LCGs are calculated from redshifts, 
obtained from strong emission lines.
     With filled black squares we also show the three
most metal-poor BCDs (SBS 0335--052E, SBS 0335--052W and I Zw 18) 
\citep{G2009}, the intermediate-redshift ($z$ $<$ 1) extremely 
low-metallicity emission-line galaxies studied by \citet{Kakazu2007}
and luminous metal-poor star-forming galaxies at $z$ $\sim$ 0.7 studied
by \citet{Hoyos2005}. 
The area occupied by the Lyman-break galaxies at $z$ $\sim$ 3
of \citet{Pettini2001} is denoted as LBG and is displayed as a 
dotted-line rectangle.
 The best linear-likelihood fit to the strongly star-forming galaxies
derived by \citet{IGT2010} 
is shown in  Fig.~\ref{LZ} by a black straight line. 
The fit to the emission-line 
galaxies studied by \citet{G2009} is shown by red line. 
We also show the location of HGs ($z <$ 0.2)
compiled by \citet{Margutti2007} by filled red circles
(980425, 030329, 031203 and 060218).
For HG 030329 and 060218 we used the metallicity 
obtained from $R_{23}$ estimates 
\citep[Table 9 in the paper of ][]{Margutti2007}
for which the metallicity calibration of \citet{Kewley2008} was applied.
Data for HGs 031203, 030329 and 060218 (green circles) are taken from  
\citet{Levesque2010}; for 020903 (light blue circles) from 
\citet{Hammer2006}, for HGs 990712, 020903, 030329, 031203 and 060218 (purple
circles) from \citet{Han2010}. Obviously, HG 031203 in the $L-M$ relation is 
placed in the region of LCGs, which
seem to form the bridge between low-mass low-metallicity BCD galaxies with 
extremely high star-forming activity and high-redshift LBGs \citep{IGT2010}.

 Thus, HG 031203 is very similar in oxygen abundance and luminosity
to LCGs and together with other HGs lies in the region 
that forms the common luminosity-metallicity 
relation, obtained by \citet{IGT2010} for low-metallicity galaxies 
with strong star-formation activity.
\citet{Watson2010} have concluded that local BCDs (in meaning, 
low-metallicity galaxies with strong 
star-formation activity) may be considered as more reliable analogues of 
star-formation in the early universe than typical local starbursts.
\citet{Levesque2010} also came to conclusion that LGRB host galaxies 
appear to fall below $L-Z$ relation for local emission-line galaxies, in the 
region occupied by metal-poor galaxies.

\subsection{Star-formation rate \label{SFR}}

One of the most 
important characteristics of galaxy evolutionary status is the 
star-formation rate SFR. We derive the SFR(H$\alpha$) 
using the extinction-corrected luminosity $L$(H$\alpha$) of the 
H$\alpha$ emission line and the relation given by \citet{Kennicutt1998},

\begin{equation}
{\rm SFR}({\rm H}\alpha)=7.9\times10^{-42} L({\rm H}\alpha). \label{eq3}
\end{equation}

In the equation, the SFR is in units of 
$M_\odot$ yr$^{-1}$, $L$(H$\alpha$) in erg s$^{-1}$, corrected for 
extinction with $C$(H$\beta$) = 1.67 at a distance of 430 Mpc 
taken from the NED.
  This results in an extinction-corrected H$\alpha$ luminosity of 
7.27 $\times$ 10$^{41}$ erg s$^{-1}$.
 The star-formation rate for HG 031203 is then
5.74 $M_\odot$ yr$^{-1}$, which is very close to other determinations, for
instance, 4.8 $M_\odot$ yr$^{-1}$, obtained by \citet{Levesque2010}.
The star-formation rate for LCGs is in 
the range 0.7 - 60 $M_\odot$ yr$^{-1}$ with an average value 
of $\sim$4 $M_\odot$ yr$^{-1}$.  
 Thus, there is no significant  
difference in the SFR for HG 031203 and average value for LCGs.
 Along the same line the range of SFRs in LCGs is comparable to that in the 
intermediate-redshift star-forming galaxies \citep{Hoyos2005,Kakazu2007} and in
LBGs \citep{Pettini2001}, but is $\sim$10-100 times higher than in 
typical nearby BCDs. For example, the SFR in 
I Zw 18 is 0.1 $M_\odot$ yr$^{-1}$ \citep{T08}.

The H$\beta$ luminosity of HG 031203 derived from the X-shooter spectrum 
and corrected for the extinction with $C$(H$\beta$) = 1.67 is 
2.55 $\times$ 10$^{41}$ erg s$^{-1}$.  
  Based on the extensive sample of LCGs
selected from SDSS DR7, \citet{IGT2010} showed that there is a
non-redshift-dependent  upper limit of  
$L$(H$\beta$) $\sim$ 2.5 $\times$ 10$^{42}$ erg~s$^{-1}$, likely due to
a self-regulating mechanism in star formation. 
All LCGs in \citet{IGT2010} range in 
$L$(H$\beta$) from 3 $\times$ 10$^{40}$ to 2.5 $\times$ 10$^{42}$ erg~s$^{-1}$.
  Thus, $L$(H$\beta$) of HG 031203 is in the range of H$\beta$ luminosities 
of strongly star-forming galaxies.

We derived the stellar mass of HG 031203 by fitting its spectrum with 
the stellar populations of different ages. This procedure was developed
by  \citet{G2006,G2007} and \citet{IGT2010}.
We include the ionised gas continuum emission into the fitting procedure 
because the neglect of this emission in actively star-forming galaxies 
with EW(H$\beta$) $\ge$ 100\AA\
leads to overestimates of galaxy stellar masses by a factor of several,
as was shown in \citet{IGT2010}.

\begin{figure}[t]
\hspace*{0.0cm}\psfig{figure=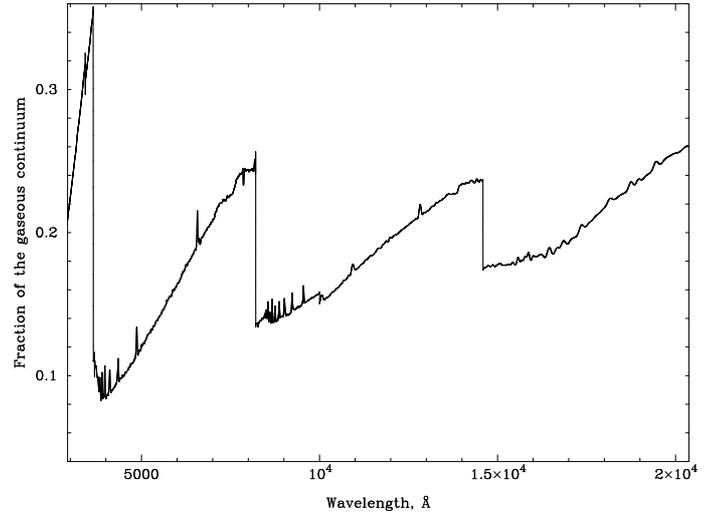,angle=-90,width=9.cm,clip=}
\caption{Fraction of gaseous 
emission to total emission vs. wavelength
for the modelled spectrum (EW(H$\beta$) = 134\AA). 
The three jumps seen at $\sim$$\lambda$3660\AA, $\sim$$\lambda$8200\AA\ 
and $\sim$$\lambda$14600\AA\ are caused by 
the hydrogen Balmer, Paschen, and Brackett discontinuities in the 
ionised gas emission.
}
\label{jumps}
\end{figure}

\begin{figure}[t]
\hspace*{0.0cm}\psfig{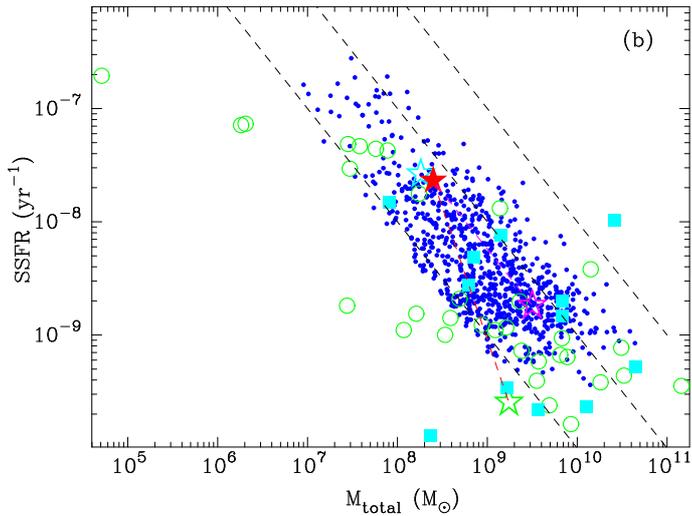}
\caption{Specific star formation rate SSFR(H$\alpha$) vs. the stellar mass
$M_*$ is shown by filled blue circles
for LCGs \citep{IGT2010}.   
The location of HG 031203 is marked by the large red star.
The dashed lines correspond from left to right to loci with SFR = 1, 10
and 100 $M_\odot$ yr$^{-1}$, respectively.
Data from \citet{Svensson2010} are shown by large green open circles.
Open green and purple stars show the position of HG 031203 from 
\citet{Svensson2010} and \citet{Watson2010}, respectively. Data 
for different HGs from 
\citet{Levesque2010b} are shown by large filled  light-blue squares 
with HG 031203 shown by a light-blue open star).
\vspace{0.1cm}
\hspace{10.5cm} (A colour version of this figure is available in the online journal.)
}
\label{ssfr}
\end{figure}

Each fit is performed over the whole observed spectral range 
 $\sim$$\lambda$$\lambda$3200 -- 24000\AA.
  The shape of the SED depends on several parameters. Because each SED is
the sum of both stellar and ionised gas emission, its shape depends
on the relative brightness of stellar and ionised gas emission.
  In strongly star-forming galaxies, the contribution of the ionised gas
can be very high.
 However, the EWs of hydrogen emission lines never attain the 
theoretical values for pure ionised gas emission. This implies a 
non-negligible contribution of stellar emission.
  We therefore parameterise the relative  contribution of gaseous 
emission to the stellar one by the equivalent width EW(H$\beta$).
  The gaseous continuum emission is calculated following \citet{Aller1984}
and includes hydrogen and helium free-bound, free-free, and two-photon 
emission.

The shape of the spectrum also depends on reddening.
  We have no direct observational constraint for the reddening 
of the stellar component, which can be different in principle
from $C$(H$\beta$) obtained from the
measured hydrogen line fluxes.

Finally, the SED depends on the star-formation history (SFH) of the galaxy.
We carried out a series of Monte Carlo simulations to reproduce the SED. 
   To calculate the contribution of stellar emission to the SEDs,
we adopted a grid of the Padua stellar evolution
models by 
\citet{Girardi2000}\footnote{http://pleiadi.pd.astro.it.}
with a heavy element mass fraction $Z$ = 0.004. 
  Using these data we calculated with the
package PEGASE.2 \citep{FR97} a grid of  
instantaneous burst SEDs in a range of ages from  0.5 Myr to 15 Gyr.
 We adopted a stellar initial mass function with a Salpeter 
slope, an
upper mass limit of 100 $M_\odot$ and a lower mass limit of 0.1 $M_\odot$.
 Then the SED with any star-formation history can be obtained by integrating
the instantaneous burst SEDs over time with a specified time-varying 
star-formation rate.
  We approximated the SFH in HG 031203 by a recent short burst, 
which accounts for the young stellar population,
and a continuous star formation responsible for the older stars.
  The contribution of each type of stellar 
population to the SED is defined by the ratio of the masses of the 
old to young stellar populations, b = $M$(old)/$M$(young), which we varied
between 0.01 and 1000.

Then the total modelled continuum flux near H$\beta$ for a mass of
1 $M_{\odot}$ is scaled to fit the extinction-corrected luminosity of the 
galaxy at the same wavelength. The scaling factor is equal to the  
stellar mass $M_*$ of the galaxy.

The flux ratio of the gaseous continuum to the total continuum depends 
on the adopted electron temperature $T_e$(H$^+$) in the H$^+$ zone, because
EW(H$\beta$) for pure gaseous emission decreases with increasing $T_e$(H$^+$).
 Given the $T_e$(H$^+$) is not necessarily equal to $T_e$(O {\sc iii}),
we chose to vary it in the range (0.7--1.3)$T_e$(O {\sc iii}).
  We assume that the extinction coefficient $C$(H$\beta$)$_{stars}$ for the 
stellar light is the same as $C$(H$\beta$)$_{gas}$ for the ionised gas, and 
vary both in the range (0.8--1.2)$C$(H$\beta$), where $C$(H$\beta$) is derived 
from Balmer decrement.

We ran 5$\times$10$^5$  Monte Carlo models 
varying simultaneously $t$(young), $t$(old), $b$, $T_e$(H$^+$),
$C$(H$\beta$)$_{gas}$ and $C$(H$\beta$)$_{stars}$.
  The best-modelled SED 
is found from $\chi ^2$ minimisation of the
deviation between the modelled and the observed continuum in ten ranges 
of the spectrum that are free of the emission lines.

In Fig.~\ref{sed} we show the best-fit model SED to the redshift- and 
extinction-corrected spectrum. 
 The contributions from the stellar and 
ionised gas components are shown by green and blue
lines, respectively. 
The sum of both stellar and ionised gas emission is shown by a red line.
 The spectrum in the whole range of wavelengths 
$\sim$$\lambda$$\lambda$3200--24000\AA\ is fitted 
quite well despite the presence of the strong telluric absorption 
features in the NIR part of the spectrum. Evidently
the contribution of gaseous emission is essential because of the 
EW(H$\beta$) = 134\AA\ for the galaxy.

The fraction of the gaseous emission
as a function of wavelength in the spectrum of the galaxy is shown in 
 Fig.~\ref{jumps}.  The three jumps at $\sim$$\lambda$3660\AA, 
$\sim$$\lambda$8200\AA\
and $\sim$$\lambda$14600\AA\ are caused by the Balmer, Paschen, and 
Brackett discontinuities of the ionised gas emission. 
  Between these jumps,
the fraction of gaseous emission increases with increasing wavelength. 
  As Fig.~\ref{jumps} shows, the fraction of ionised gas 
continuum-emission in the galaxy with EW(H$\beta$) = 134\AA\ increases
from $<$10\% to $\sim$25\% (in the wavelength range $\sim$4000--8000\AA), 
from $\sim$14\% to $\sim$24\% (in the wavelength range 
$\sim$9000--15000\AA) and from
$\sim$18\% to $\sim$26\% (in the wavelength range $\sim$15000--20000\AA).

The model stellar  SED 
shown in Fig.~\ref{sed} by the green line 
is used for the stellar mass ($M_*$) determination. 
 The mass of HG 031203 is $M_*$ = 2.5$\times$10$^8$$M_{\odot}$,
which is characteristic for a dwarf galaxy.
Then the specific star-formation rate (SSFR) which is defined as 
SSFR(H$\alpha$) = SFR(H$\alpha$)/$M_*$ for the galaxy, is
2.3$\times$10$^{-8}$ yr$^{-1}$. 
  In Fig.~\ref{ssfr} the SSFR(H$\alpha$) vs. 
the stellar masses $M_*$ for LCGs \citep{IGT2010} 
are shown by filled blue circles. They vary in the range 
$\sim$10$^{-9}$--10$^{-7}$ yr$^{-1}$. 
 These values are extremely high  and similar to those found 
in high-redshift galaxies at $z$ = 4--6 \citep{Stark2009}  and $z$ = 6--8 
\citep{Schaerer2010}. The position of HG 031203 is 
denoted by a large red star.
The dashed lines correspond from left to right to loci with SFR = 1, 10
and 100 $M_\odot$ yr$^{-1}$, respectively.
\citet{Svensson2010} obtained the properties (stellar masses, 
SFRs and SSFRs) for 34 HGs, using a large multi-wavelength 
(0.45--24$\mu$) dataset from GOODS and PANS surveys.  
Data from \citet{Svensson2010} are shown Fig.~\ref{ssfr} by large green 
open circles. 
  Using an unprecedentedly wide wavelength range 
($\sim$0.3--2$\times$10$^5$$\mu$m), \citet{Watson2010} have estimated a stellar 
mass of HG 031203 as log($M_*$/$M_{\odot}$) $\sim$ 9.5. 
With our estimation of SFR(H$\alpha$) we denote the position of HG 031203 
in Fig.~\ref{ssfr} (using stellar mass of \citet{Watson2010})
by open purple star.

There is an offset between the HG sample and LCGs to higher mass
(Fig.~\ref{ssfr}) that can be explained by overestimation of the galaxy's 
stellar mass if the ionised gas continuum emission is not included in the 
fitting of the SED. Several HGs have lower SFRs (those in the left lower 
part of Fig.~\ref{ssfr}). In these cases LGRBs arise in ordinary 
emission-line galaxies with moderate star-formation activity. 
For comparison, the SFR in the prototype BCD I Zw 18 is much lower, 
0.1 $M_{\odot}$ yr$^{-1}$ \citep{T08}.

It is interesting to note that all parameters of the HG 031203 host 
galaxy and other HGs
put them into the class of LCGs. It is reasonable to adopt that many LGRBs 
occured in the most luminous compact galaxies with low metallicities, because 
they are characterised by the most vigorous star formation.
  Overall, we find no appreciable difference in heavy element abundances, 
elements ratios, luminosities, star-formation rates and specific 
star-formation rates between HG 031203 and other HGs and LCGs, implying 
their similar nature.

\section{Conclusions \label{concl}}

We have studied the spectrum of the GRB 031203 host galaxy 
(HG 031203) with the VLT/X-shooter spectroscopic observations in the 
wavelength range $\sim$$\lambda$$\lambda$3200 -- 24000\AA. These data were 
compared with the data obtained previously by \citet{IGT2010}
for luminous compact emission-line galaxies (LCGs) from SDSS DR7.
 We have arrived at the following conclusions:

1. We derive the oxygen abundance 
of 12 + log O/H = 8.20 $\pm$ 0.03 in the H {{\sc ii}} region of HG 031203.
The previous direct determinations 
\citep{Prochaska2004,Hammer2006,Margutti2007,Wiersema2007,Levesque2010,Han2010}
give the oxygen abundance 12 + log O/H in the range $\sim$ 8.0 -- 8.2.

2. We find that the extinction-corrected fluxes of hydrogen Balmer,  
Paschen, and Brackett lines 
agree well with the theoretical recombination values 
if a single value
of the extinction coefficient $C$(H$\beta$) = 1.67
is adopted. This implies that there is no  
additional star formation
that is seen in the NIR range but is hidden in the visible range.  
The star-forming region observed in the optical range is the only source of 
ionisation.

3. Using {\sl Spitzer} MIR emission-line fluxes, we have 
also found that MIR data do not reveal an additional star formation that is 
hidden at shorter wavelengths.
Thus, the emission-line spectrum of  HG 031203 in the whole 
$\sim$ 0.36 -- 20 $\mu$m wavelength range originates in relatively transparent 
H {\sc ii} regions.

4. The profiles of strong emission lines are decomposed 
into two Gaussian narrow and broad components  with
FWHM $\sim$115 and $\sim$270 km s$^{-1}$, respectively. 
These components, separated by $\sim$34 km s$^{-1}$,
likely correspond to two H {\sc ii} regions with different extinction, which 
is larger in the region with a broad component.

5. We derive a stellar mass $M_*$ = 2.5$\times$10$^8$$M_{\odot}$ 
for HG 031203 by fitting its spectrum with 
the stellar populations of different ages \citep{IGT2010,G2006,G2007}. 
We include the ionised gas continuum emission in the fitting procedure 
because the neglect of this emission in the actively star-forming galaxy 
with EW(H$\beta$)=134\AA\ leads to an overestimate of the $M_*$ 
\citep{IGT2010}.

6. We find that the heavy element abundances, element abundance ratios, 
extinction-corrected H$\alpha$ luminosity  
$L$(H$\alpha$)=7.27 $\times$ 10$^{41}$ erg s$^{-1}$, star-formation 
rate SFR(H$\alpha$) = 5.74 $M_\odot$ yr$^{-1}$ and 
specific star-formation rate
SSFR(H$\alpha$) = SFR(H$\alpha$)/$M_*$ = 2.3$\times$10$^{-8}$ yr$^{-1}$ 
of HG 031203 and other HGs are in the range occupied by
the LCGs from SDSS DR7 studied by \citet{IGT2010}.
In [O {\sc iii}] $\lambda$5007/H$\beta$ vs.
[N {\sc ii}] $\lambda$6583/H$\alpha$ diagnostic diagram \citep{BPT1981} and
in the luminosity-metallicity diagram the GRB host galaxy is also placed  
in the region of the LCGs.
 This implies that LCGs  with extreme star formation 
\citep[containing also green pea galaxies as a subclass 
of LCGs][]{Cardamone2009} may predominantly harbour the long-duration GRBs.

\acknowledgements
N.G.G., Y.I.I. and K.J.F. are grateful to the staff of the Max Planck 
Institute for Radioastronomy for their warm hospitality and 
acknowledge support through DFG grant No. FR 325/59-1. 
This research has made use of the NASA/IPAC Extragalactic Database (NED), 
which is operated by the Jet Propulsion 
Laboratory, California Institute of Technology, under contract with the 
National Aeronautics and Space Administration.




\end{document}